\documentclass[onecolumn]{aastex6}

\newcommand{\tco}{$^{13}$CO}
\newcommand{\co}{$^{12}$CO}

\newcommand{\tcoone}{$^{13}$CO $J$=1$-$0}
\newcommand{\coone}{$^{12}$CO $J$=1$-$0}

\newcommand{\cotwo}{$^{12}$CO $J$=2$-$1}
\newcommand{\tcotwo}{$^{13}$CO $J$=2$-$1}

\newcommand{\cothree}{$^{12}$CO $J$=3$-$2}

\newcommand{\lfir}{$L_{\rm{FIR}}$}
\newcommand{\lsol}{$L_{\odot}$}
\newcommand{\msol}{$M_{\odot}$}
\newcommand{\mmol}{$M_{\rm{H_{2}}}$}
\newcommand{\kms}{km s$^{-1}$}
\newcommand{\tkin}{$T_{\rm{kin}}$} 
\newcommand{\nhtwo}{$n_{\rm{H_{2}}}$}
\newcommand{\alphacou}{$M_{\odot}$ (K km s$^{-1}$ pc$^{2}$)$^{-1}$}
\newcommand{\alphaco}{$\alpha_{\rm{CO}}$}

\newcommand{\nco}{$N_{\rm{^{12}CO}}$}
\newcommand{\ff}{$\Phi_{\rm{A}}$}

\newcommand{\xco}{[$^{12}$CO]/[$^{13}$CO]}

\newcommand{\xh}{$x_{\rm{CO}}$}
\newcommand{\arc}{$^{\prime\prime}$}

\shorttitle{Luminous Infrared Galaxies with the SMA V}
\shortauthors{Sliwa et al.}
\begin{document}


\title{Luminous Infrared Galaxies with the Submillimeter Array V: Molecular Gas in 
Intermediate to Late-Stage Mergers}

%% Use \author, \affil, plus the \and command to format author and affiliation 
%% information.  If done correctly the peer review system will be able to
%% automatically put the author and affiliation information from the manuscript
%% and save the corresponding author the trouble of entering it by hand.
%%
%% The \affil should be used to document primary affiliations and the
%% \altaffil should be used for secondary affiliations, titles, or email.

%% Authors with the same affiliation can be grouped in a single
%\author{Greg J. Schwarz\altaffilmark{1,2} and August Muench\altaffilmark{1}}
%\affil{American Astronomical Society \\
%2000 Florida Ave., NW, Suite 300 \\
%Washington, DC 20009-1231, USA}

\author{Kazimierz Sliwa\altaffilmark{1,2}, Christine D. Wilson\altaffilmark{2},  Satoki Matsushita\altaffilmark{3},  Alison B. Peck\altaffilmark{4}, Glen R. Petitpas\altaffilmark{5}, Toshiki Saito\altaffilmark{6,7} \and Min Yun\altaffilmark{8}}
\altaffiltext{1}{Max Planck Institute for Astronomy, K{\"o}nigstuhl 17, D-69117 Heidelberg, Germany; sliwa@mpia-hd.mpg.de}
\altaffiltext{2}{Department of Physics and Astronomy, McMaster University, Hamilton,  ON L8S 4M1 Canada; wilson@physics.mcmaster.ca}
\altaffiltext{3}{Academia Sinica Institute of Astronomy and Astrophysics, P.O. Box 23-141, Taipei 10617, Taiwan}
\altaffiltext{4}{Gemini Observatory, Northern Operations Center, 670 N. A'ohoku Pl, Hilo, HI 96720}
\altaffiltext{5}{Harvard-Smithsonian Center for Astrophysics, Cambridge, MA 02138}
\altaffiltext{6}{Department of Astronomy, The University of Tokyo, 7-3-1 Hongo, Bunkyo-ku, Tokyo 113-0033, Japan}
\altaffiltext{7}{National Astronomical Observatory of Japan, 2-21-1 Osawa, Mitaka, Tokyo 181-8588, Japan}
\altaffiltext{8}{Department of Astronomy, University of Massachusetts, Amherst, MA 01003, USA}
%% Use the \and command so offset the last author.
%\and

%\author{Jeff Lewandowski\altaffilmark{5}}
%\affil{IOP Publishing, Washington, DC 20005}


%\altaffiltext{1}{AAS Journals Data Scientist}

\begin{abstract}

We present new high-resolution ALMA (\tcoone\ and $J$= 2-1) and CARMA (\co\ and \tcoone) observations of two 
Luminous Infrared Galaxies (LIRGs): Arp~55 and NGC~2623. The new data are complementary to  
published and archival Submillimeter Array observations of \cotwo\ and $J$=3--2. We 
perform a Bayesian likelihood non-local thermodynamic equilibrium analysis to constrain the molecular gas 
physical conditions such as temperature, column and volume densities and the \xco\ abundance ratio. For Arp~55, an early/intermediate staged merger, the line measurements are consistent with cold ($\sim$ 10-20~K), dense ($>$ 10$^{3.5}$ cm$^{-3}$) molecular gas.
For NGC~2623, the molecular gas is warmer ($\sim$ 110~K) 
and less dense ($\sim$ 10$^{2.7}$ cm$^{-3}$). Since Arp~55 is an early/intermediate 
stage merger while NGC~2623 is a merger remnant, the difference in physical conditions 
may be an indicator of merger stage. Comparing the temperature and volume density of several 
LIRGs shows that the molecular gas, averaged over $\sim$kiloparsec scale, of advanced mergers is in general warmer and less 
dense than early/intermediate stage mergers. We also find that the \xco\ abundance ratio 
of NGC~2623 is unusually high ($>$250) when compared to the Milky Way; however, it follows a trend seen with other LIRGs in literature. This high \xco\ value is very likely due to stellar nucleosynthesis enrichment of the interstellar medium. On the other hand, Arp~55 has a more Galactic \xco\ value with the most probable \xco\ value being 20-30. We measure the CO-to-H$_{2}$ conversion factor, \alphaco, to be $\sim$ 0.1 and $\sim$ 0.7 (3 $\times$ 10$^{-4}$/\xh) \alphacou\ for Arp~55 and NGC~2623, respectively. Since Arp 55 is an early/intermediate , this suggests that the transition from a Galactic conversion factor to a LIRG value happens at an even earlier merger stage. 

\end{abstract}




\section{Introduction} \label{sec:intro}

Galaxy mergers are an important process in the evolution of galaxies, as shown by 
the observed increase in merger rate in the early universe \citep[e.g.][]{Bridge2010}. The 
gravitational interactions between the galaxies significantly alter the morphology, 
luminosity, colour, size, star formation rate, and mass distribution in a relatively short 
period of time. In particular, the molecular gas, which fuels the current and future star formation, begins in a structured morphology in the progenitor galaxies \citep[assuming disk galaxies as the progenitors see M51 for an example;][]{Pety2013,Schinnerer2013}. During the merger event, the gas is driven towards the inner few kiloparsec regions triggering an intense starburst and/or feeding a supermassive 
black hole \citep[e.g.][]{Hopkins2006}. The ``classical" scenario for the merger 
process leads to the formation of an elliptical galaxy \citep{Toomre1977}; however, there is 
observational evidence that mergers can reemerge as disk galaxies 
\citep[e.g.][]{Ueda2014}. Nevertheless, the merger process is still not well understood. We 
still do not have a physical connection between the merger process and star formation and in 
particular, the timescales and processes controlling star formation in mergers. High-resolution observational 
studies of Ultra/Luminous Infrared Galaxies (U/LIRGs) at different stages of the merger are needed 
to further our knowledge of what happens to the molecular gas during the merger.

%Molecular gas in external galaxies has been best traced by the most common isotopologue of carbon monoxide,\co. 
An approach to understanding the physical conditions of the molecular gas during the merger process is to use a radiative transfer method. This method requires multiple transitions of molecular lines which were difficult to obtain a decade ago. Several starburst galaxies observed with \textit{Herschel}'s Fourier Transform Spectrometer (FTS), which could observe several \co\ transitions simultaneously, were modelled using a non-local thermodynamic equilibrium (non-LTE) method \citep[e.g.][]{Panuzzo2010,Rangwala2011,Kamenetzky2012,Kamenetzky2014,Schirm2014,Tunnard2015b}. The FTS spectral line energy distributions (SLEDs) of the modeled starbursts are consistent with a two-component molecular ISM: cold, moderately dense gas and a hot, very dense gas. The high-$J$ transitions ($\geq$ 4-3) are important to constrain the hot molecular gas component properties, lines that are difficult to obtain from the ground with only a few local galaxies observed in such transitions \citep[e.g.][]{Sliwa2013,GarciaBurillo2014,Xu2014,Xu2015,Rangwala2015,Krips2016,Aalto2017}. Modeling the molecular gas at higher resolution has been difficult with only a few starbursts analyzed (e.g. \citealt{Matsushita2009,Sliwa2012,Sliwa2013,Sliwa2014,Saito2017b}; Sliwa $\&$ Downes submitted) due in part to the lack of high-resolution observations of optically thin tracers such as  \tco. However, with the Atacama Large Millimeter/submillimeter Array (ALMA) and the Northern Extended Millimeter Array (NOEMA; formerly the Plateau de Bure Interferometer), observations of such weakly emitting tracers have become routine. Eventually with ALMA, the higher $J$ CO transitions will be observed and we will start to probe the hot molecular gas components that \textit{Herschel} has observed.

In this paper, we present new high-resolution ALMA (\tcoone\ and $J$=2-1) and Combined Array for Research in Millimeter-wave Astronomy (CARMA; \co\ and \tcoone) observations for two LIRGs: Arp~55 and NGC~2623. These LIRGs are at very different merger stages and offer snapshots at how the molecular gas evolves over the merger process. The new observations complement existing Submillimeter Array (SMA) data and we use the measured line intensities to perform a non-LTE analysis to determine the molecular gas physical conditions on $\sim$1~kpc scales. The paper is organized as follows: In Section \ref{sec:obs}, we present the targets and describe the observations, data reduction and short spacings details. In Section \ref{sec:morphology}, we describe the new \coone\ observations from CARMA, where we detect new features, compared to the previously published maps. In Section \ref{sec:moleculargas}, we present the non-LTE modeling and results. In Section \ref{sec:discussion}, we discuss the results of the non-LTE modeling such as the \xco\ abundance ratio, the \coone\ luminosity to mass conversion factor (\alphaco) and the evolution of the molecular gas as a function of merger stage on kiloparsec scales. In Appendix \ref{sec:lineratio}, we present the various integrated brightness temperature line ratio maps. In Appendix \ref{sec:velmaps}, we present the velocity field and dispersion ($\sigma$ = FWHM/2$\sqrt{2\rm{ln}(2)}$) maps from the CARMA observations.

\section{Observations} \label{sec:obs}    %%%%%%%%%%%%
%OBSERVATIONS
In the sections below, we present the targets and then describe the observations in detail. Table 
\ref{tab:summary} summaries the properties of the observations. Table \ref{tab:fluxdata}
presents the measured line fluxes for each LIRG and the molecular gas mass measured from the 
\coone\ line assuming a standard U/LIRG conversion factor of 0.8 \alphacou\ \citep{Downes1998}.

\subsection{The Targets} \label{sec:targets}
\objectname{Arp~55} (UGC~4881, VV~115, IRAS~09126+4432)  is one of the least 
studied LIRGs (\lfir\ = 4.0 $\times$ 10$^{11}$ \lsol; \citealt{Sanders2003}) in the sample of \cite{Wilson2008} (Paper I). The 
system consists of two nuclei separated by $\sim$ 11\arc\ (9.3~kpc at $D_{\rm{L}}$ = 
175~Mpc; 1\arc\ = 850~pc). A tidal tail is seen in the optical extending from the 
north-eastern galaxy that is also visible in H$\alpha$ \citep{Hattori2004}. Four H$\alpha$ 
emission sources are detected in the tail. The south-western galaxy has compact H$\alpha
$ emission with a slight extension in the north-south direction. \cite{Vardoulaki2015}, using Very Large Array (VLA) radio continuum, classify the north-eastern nucleus as a composite AGN/starburst system and the south-western nucleus as a radio-AGN. The stellar mass, 
determined from photometry, is $\sim$ 7 $\times$ 10$^{10}$ \msol\ \citep{U2012}. 
The total star formation rate is $\sim$ 60 \msol\ yr$^{-1}$ \citep{U2012} determined 
using UV and infrared emission. Interferometric observations of \coone\ revealed 
molecular gas in the region containing the two nuclei \citep{Sanders1988}. 

\objectname{NGC~2623} (Arp~243, VV~79, IRAS~08354+2555) is an advanced 
merger LIRG (\lfir\ = 3.0 $\times$ 10$^{11}$ \lsol; \citealt{Sanders2003}) at a distance of 83~Mpc (1\arcsec\ = 402~pc). \cite{Privon2013} show 
via numerical simulations that NGC 2623 is a merger remnant with the two nuclei 
coalescing roughly 80 Myrs ago. The system has two prominent tidal tails that 
extend 20--25\arc\ to the northeast and southwest \citep{Evans2008}. X-ray 
observations show evidence of a Compton-thick AGN \citep{Maiolino2003}. 
\cite{Imanishi2009} and \cite{Privon2015} showed that the HCN/HCO$^{+}$ $J$=1--0 line ratio is above 1, 
consistent with a system that contains an AGN; however, we do note that the HCN/HCO$^{+}$ diagnostic is still under debate as starburst and composite systems may have ratios similar to AGN systems \citep[see][]{Privon2015}. The total star formation rate is 
measured to be $\sim$ 42 \msol\ yr$^{-1}$ \citep{U2012} with a stellar mass of $
\sim$ 5.5$\times$ 10$^{10}$ \msol\ \citep{Gonzalez-Delgado2014}. Near and mid-infrared 
observations reveal a compact nucleus \citep{Scoville2000, Soifer2001}. Some of 
the first high-resolution \coone\ observations also showed a compact nucleus 
\citep{Bryant1999}.

\begin{deluxetable*}{ccccccc} %%%TABLE 1
\tablewidth{0pt}
\tablecaption{Observations Summary \label{tab:summary}} 
\tablehead{\colhead{Source} & \colhead{Emission} &  \colhead{Observatory} &  \colhead{LAS
\tablenotemark{a}} &  \colhead{rms} &  \colhead{Channel Width} &  \colhead{Beam
\tablenotemark{b}}   \\ 
 \colhead{} & \colhead{} &  \colhead{} &  \colhead{(arcsec)} &  \colhead{(mJy beam$^{-1}$)} 
&  \colhead{(\kms)} & \colhead{(arcsec)}   }
\startdata
Arp 55 	& \coone\ 		& CARMA 	& 41.8 	& 3.0 	& 20 	& \textbf{1.8 $\times$ 1.7}  
\\
		& \cotwo\ 		& SMA  		& 12 		&12.6 	& 30	& 1.0 $\times$ 0.8  \\
		& \cothree\ 	& SMA  		& 4.1  	& 29 		& 25 	& 0.9 $\times$ 0.8   \\
		& \tcoone\  	& CARMA  	& 43.8 	 & 1.1	& 100  	& 1.7 $\times$ 1.6 \\
		& \tcotwo\ 		& SMA  		& 12.5 	& 8.1		& 50  	& 3.9 $\times$ 2.5 \\
		& 3mm  		&CARMA		&41.8	&0.18	&--		&2.6 $\times$ 2.3 	
%Arp 299 	& \coone\ 		& CARMA 	& 41.8 	& 2.7 	& 20 	& \textbf{1.8 $\times$ 1.7}  
\\
%		&			&		&		&	&	\\
NGC 2623 & \coone\ 	& CARMA 	& 27.3 	 &4.6		 &	 20 & 2.5 $\times$ 1.7 \\
		& \cotwo\ 		& SMA  		& 11.7 	&21.1 	& 20 & 2.1 $\times$ 1.7 \\
		& \cothree\ 	& SMA  		& 9.9 	&13.6		& 20 & 2.2 $\times$ 2.0  \\
		& \tcoone\  	& ALMA  		&28.6  	&0.43 		&30 & 2.4 $\times$ 1.9    \\
		& \tcotwo\		& ALMA		&14.3	&1.1		&40 & 3.4 $\times$ 1.4 \\
		& 3mm  		&CARMA		&28.6 	&0.23	&--	&	2.9 $\times$ 2.2 \\
\enddata
\tablenotetext{a}{The largest angular scale is recovered determined by the shortest 
projected baseline: 0.6$\lambda$/d$_{\rm{min}}$}
\tablenotetext{b}{The resolution used for the analysis is bolded. For NGC 2623, we use a compromise angular resolution of 2.5\arc. }
\end{deluxetable*}

\subsection{ALMA} \label{sec:ALMA}

ALMA observations towards NGC~2623 were conducted during Cycle~2 (Project ID= 2015.1.00804.S; PI: K.Sliwa). The receiver Bands 3 and 6 were tuned to \tcoone\ and $J$ = 2-1, respectively. Each observation consisted of four spectral windows with a bandwidth of 1.875~GHz and 960 channels each. The Band 3 data were calibrated using the \verb=CASA= \citep{McMullin2007} ALMA pipeline (v4.5.3) and the calibration solutions were checked for any obvious problems. The observations used J0854+2006 as the flux and bandpass calibrators and J0837+2454 as the gain calibrator. The Band 6 observations were manually calibrated using \verb=CASA= v4.7.0. The data were observed during an array configuration transition and contain baselines up to 6~km; however, the $uv$-coverage is poor for the long baselines and we flag all baselines above 200~m after applying the water vapour radiometer and system temperature corrections. The calibration was done in standard fashion using J0854+2006 as the flux and bandpass calibrator and J0830+2410 as the gain calibrator. We estimate a flux calibration uncertainty of 10$\%$ and 20$\%$ for Bands 3 and 6, respectively. 

Each dataset was continuum subtracted in the $uv$ domain using line-free channels. We image the datasets using \verb=CASA= v4.7.0 creating datacubes of 30 and 40 \kms\ channel width for \tcoone\ and $J$=2-1, respectively, using a Briggs robust weighting of 0.5. CLEAN boxes were placed around the emission to aid the CLEAN process converge. Integrated intensity maps were created including line emission channels and using a 2$\sigma$ cutoff. Each integrated intensity map was primary beam corrected. 

Other lines and continuum present in the ALMA datasets will be discussed in a forthcoming paper (Sliwa et al. in prep.).

\subsection{CARMA} 
\label{sec:CARMA}

Arp 55 was observed in \co\ and \tcoone\ using CARMA in the B, C, and D array 
configurations on 15 January 2013, 06 December 2012 and 18 April 2012, respectively. 
NGC 2623 was observed in  \co\  using CARMA in the  C and D array 
configurations on 2 April 2012 and 2 May 2012, respectively. We convert the data to 
\verb=CASA= measurement sets for calibration and imaging. We use 3C84 and Mars as 
the flux calibrators and 0927+390, 0854+201 and 0920+446 as the gain and bandpass 
calibrators. For Mars, we use the `Butler-JPL-Horizons 2012' standard model to compute the flux. Imaging was done in a similar manner to the ALMA observations (see Section \ref{sec:ALMA} and Figures \ref{fig:arp55maps} and \ref{fig:ngc2623maps}).

We combined the LSB and USB line-free spectral windows using a `multi-frequency synthesis' (\verb=mfs= mode in \verb=CASA=) approach to image the 2.8~mm continuum. The data were CLEANed down to the 2$\sigma$ level with the help of a clean box around the emission. For Arp 55, we degrade the resolution (see Table \ref{tab:fluxdata}) to increase the surface brightness sensitivity. We primary-beam corrected the maps (Figure \ref{fig:arp55maps}, \ref{fig:ngc2623maps}) and applied a 2$\sigma$ cutoff to measure the continuum. 

We measure the 2.8~mm continuum flux of NGC 2623 and Arp 55 to be 18 $\pm$ 4 mJy and 1.9 $\pm$ 0.4 mJy, respectively. For Arp 55, the north-eastern nucleus has a flux of 0.8 $\pm$ 0.2 mJy and the south-western nucleus has a flux of 1.1 $\times$ 0.2 mJy. See Table \ref{tab:summary} for details on the images.

\subsection{Submillimeter Array (SMA)} \label{sec:SMA}
We use the SMA calibrated $uv$ datasets of \cotwo\ and $J$ = 3--2 from Paper I. For 
calibration details, see Paper I. Arp 55 was also observed using the very extended (PI: V. 
U) and compact (PI: Z. Wang) configurations in \cotwo. The \tcotwo\ line is also observed 
simultaneously in the lower sideband. We obtained the raw datasets from the archive 
and converted them to \verb=CASA= measurement sets in order to 
calibrate and image the data. Calibration was done using the standard method using  
Callisto as the flux calibrator and 0927+390, 0920+446, and 3C84 as the gain and 
bandpass calibrators. For Callisto, we use the `Butler-JPL-Horizons 2012' standard model to compute the flux.

\begin{figure*}[htbp] %%%FIGURE 1 Arp 55 maps
\centering

\gridline{\fig{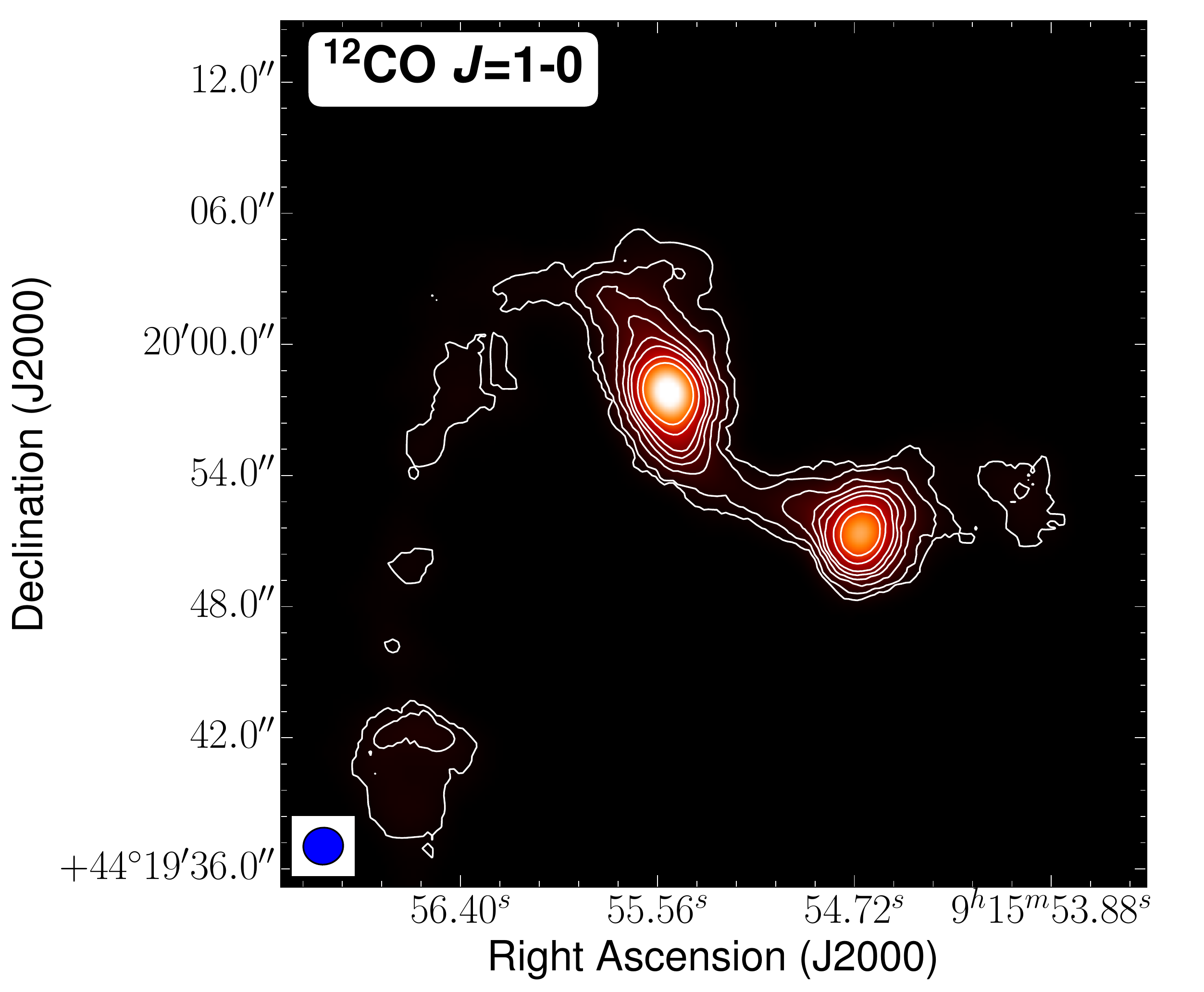}{0.4\textwidth}{(a)}
 \fig{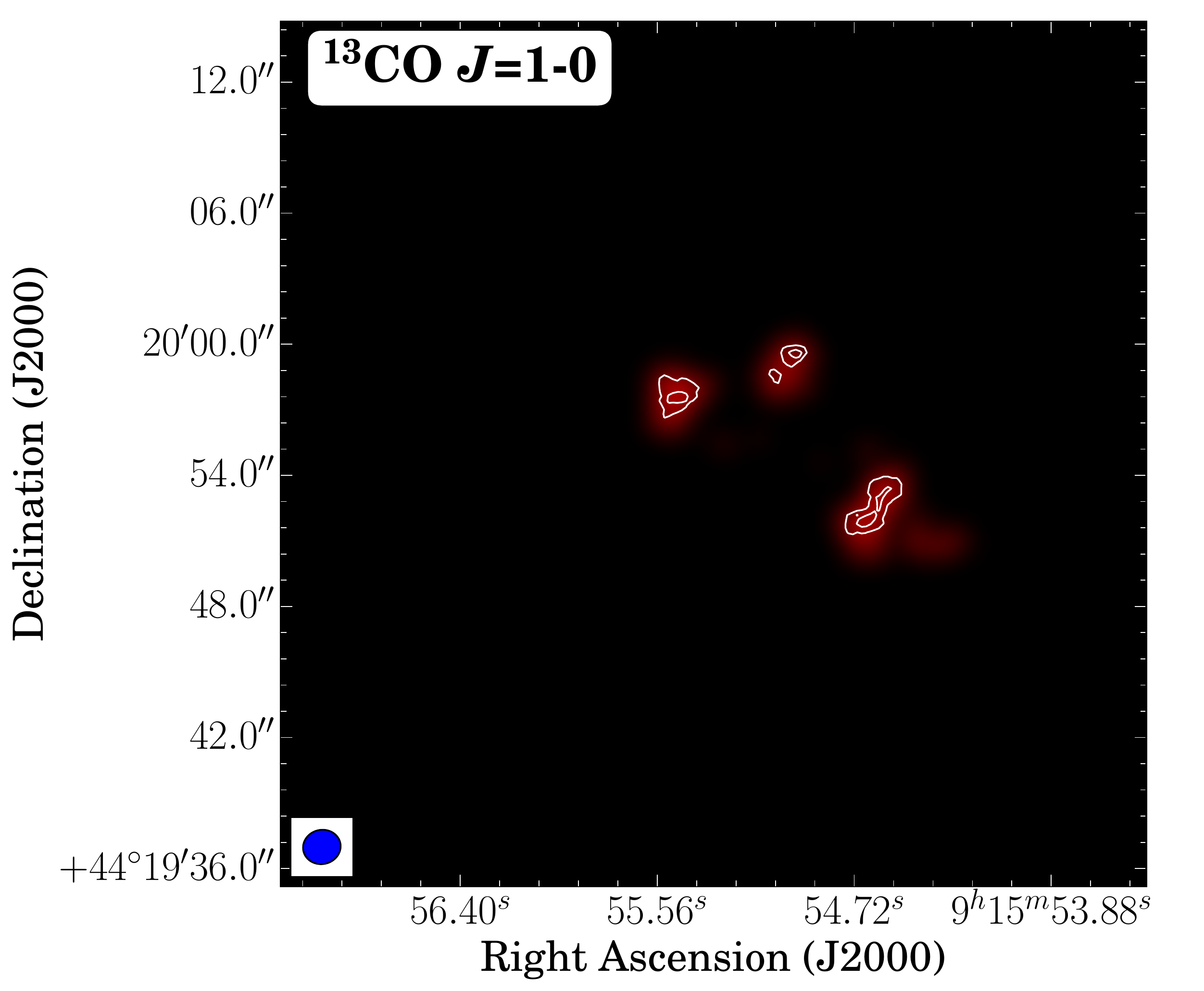}{0.4\textwidth}{(b)} }
 
\gridline{\fig{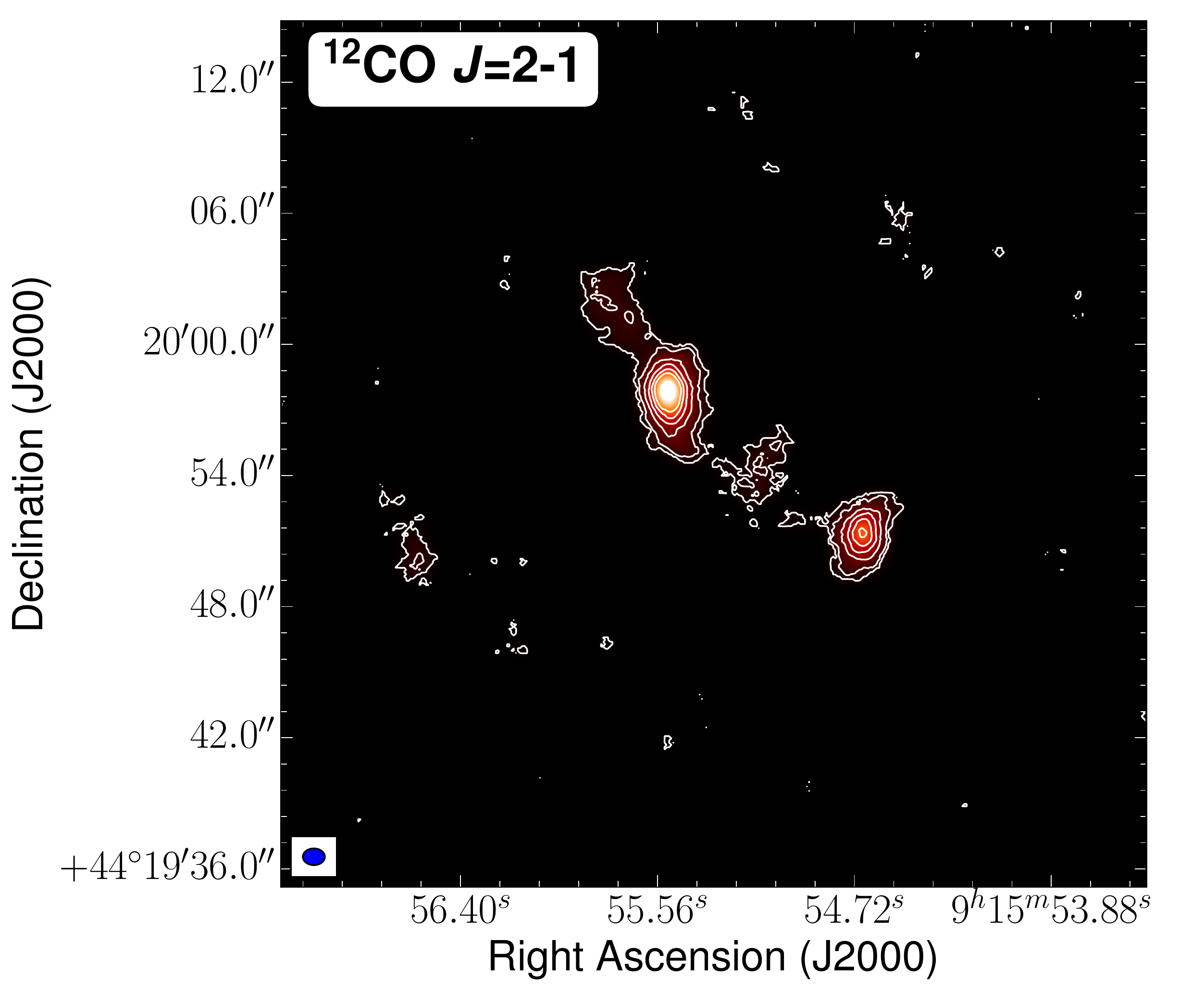}{0.4\textwidth}{(c)} 
\fig{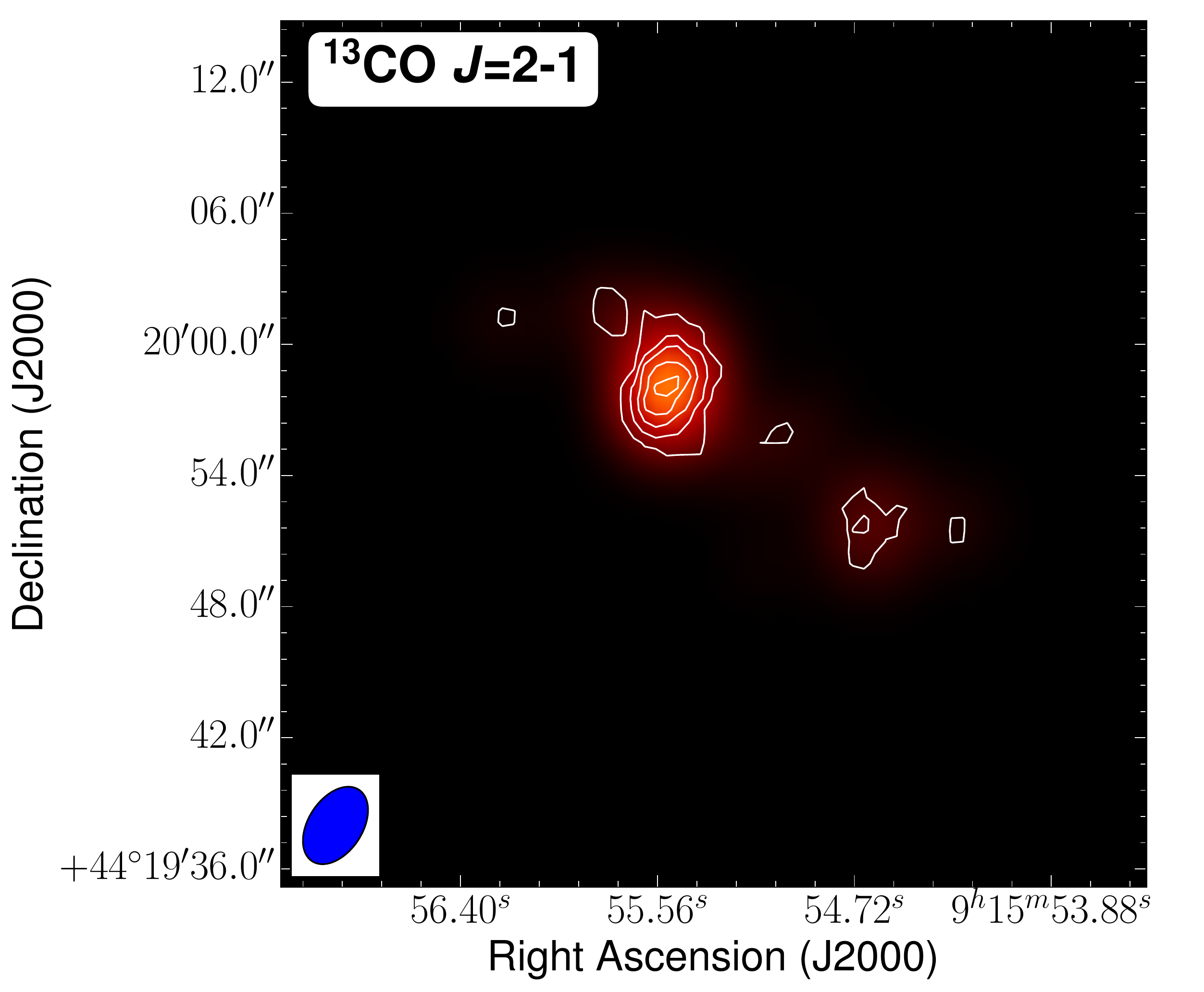}{0.4\textwidth}{(d)}} 

\gridline{\fig{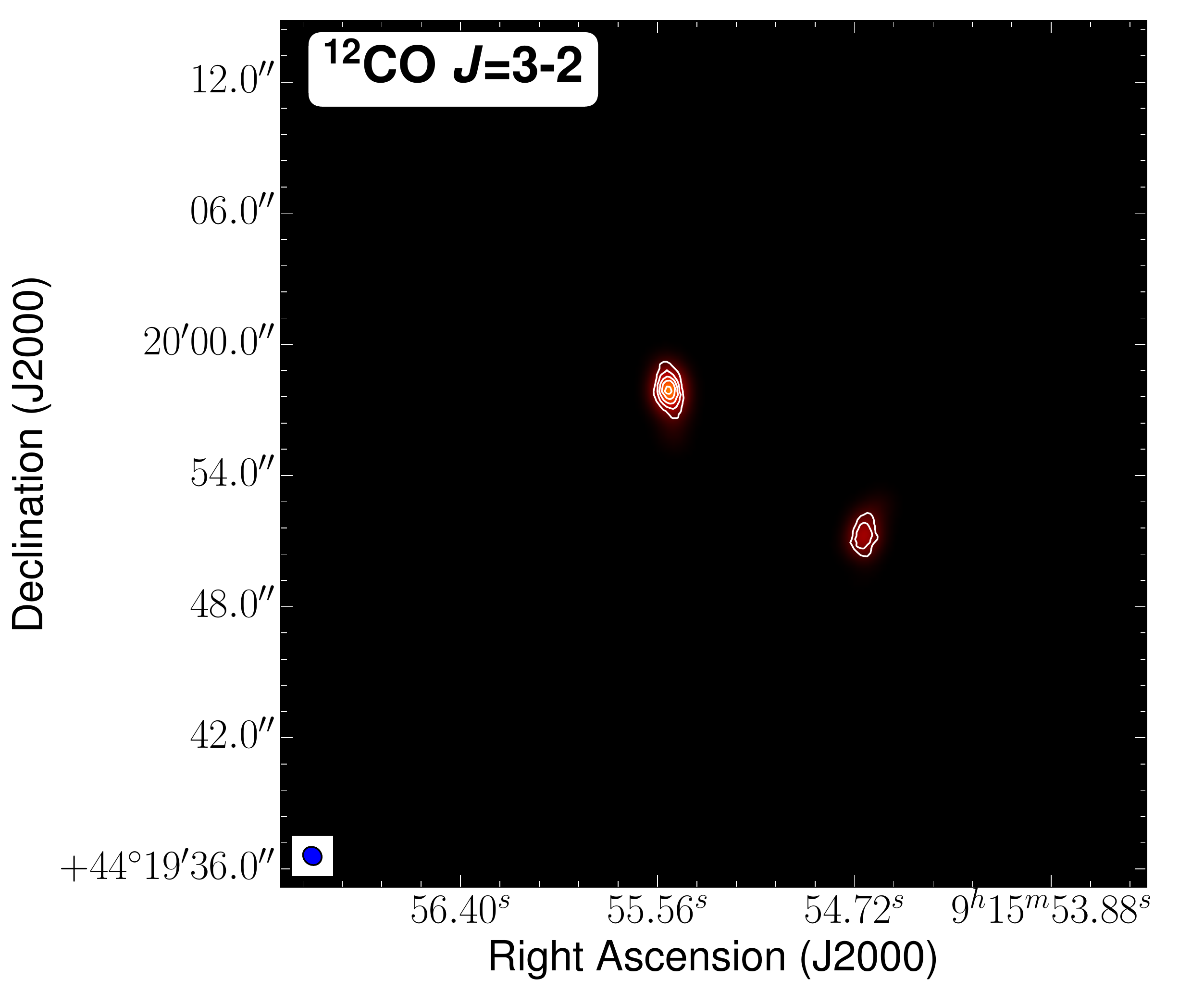}{0.4\textwidth}{(e)} 
\fig{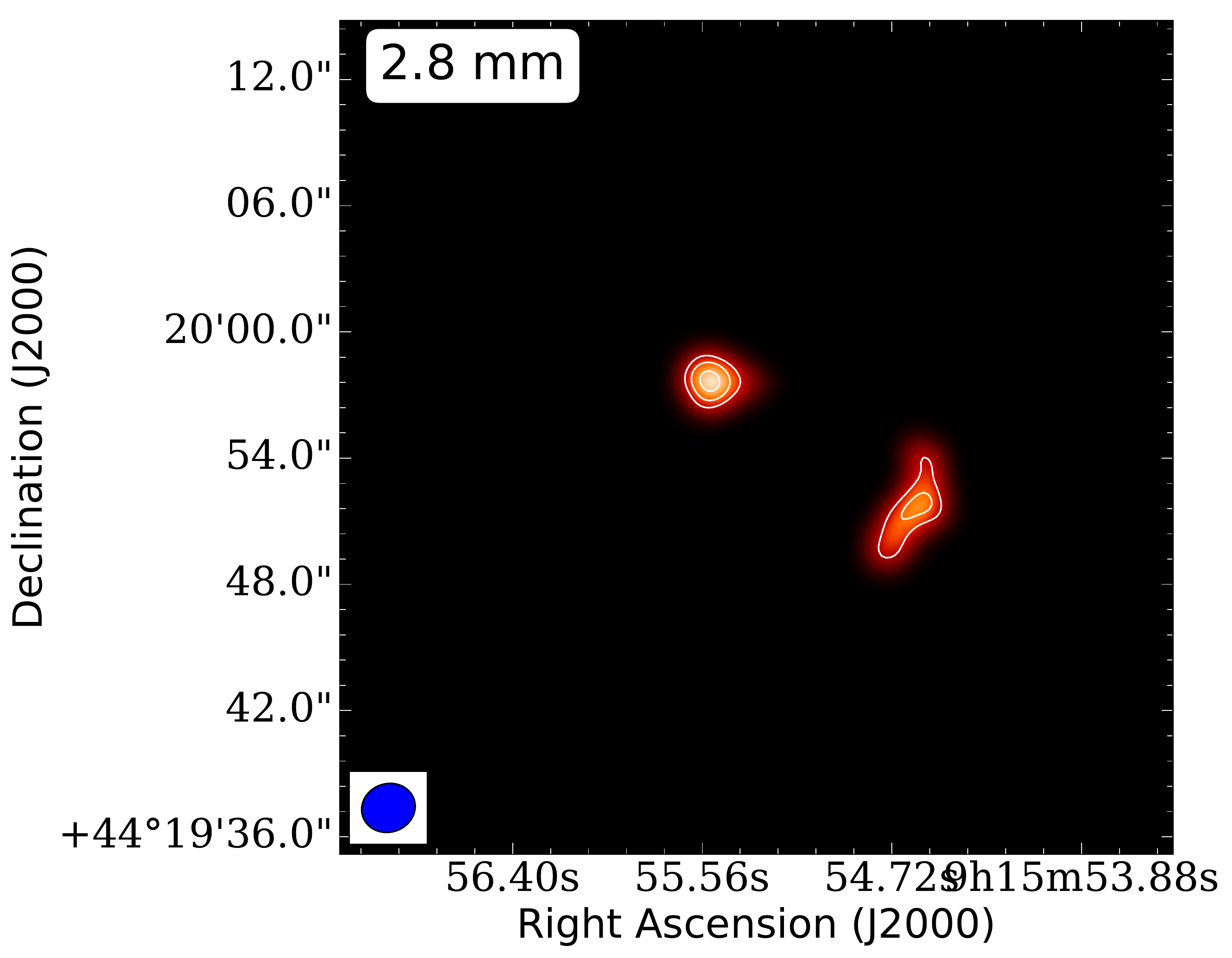}{0.4\textwidth}{(f)}} 

\caption{Arp 55: (a) CARMA \coone\ map. Contours correspond to (1, 2, 4, 6, 8, 
10, 15, 20) $\times$ 2$\sigma$ (= 0.56 Jy beam$^{-1}$ \kms).(b) CARMA \tcoone\ map. Contours 
correspond to (1.5, 2, 2.5) $\times$ 2$\sigma$ (= 0.44 Jy beam$^{-1}$ \kms). 
(c) SMA \cotwo\ map. Contours correspond to (1, 2, 4, 6, 8, 10, 15, 20) $\times
$ 1$\sigma$ (= 3.5 Jy beam$^{-1}$ \kms). (d) SMA \tcotwo\ map. Contours correspond to (1,2, 3, 4, 5) 
$\times$ 1$\sigma$ (= 1.8 Jy beam$^{-1}$ \kms).
(e) SMA \cothree\ map. Contours correspond to (2, 4, 6, 8, 10) $\times$ 1$\sigma$ 
(= 7.0 Jy beam$^{-1}$ \kms). (f) CARMA 2.8~mm continuum map. Contours correspond to (3,4,5) $\times$ 1$\sigma$ (= 0.18 mJy beam$^{-1}$).
The ellipse in the bottom left corner of each map represents the synthesized beam. All maps 
have been primary beam corrected. \label{fig:arp55maps} }
\end{figure*}

\begin{figure*}[htbp] %%%FIGURE 2 NGC 2623 maps
\centering
\gridline{\fig{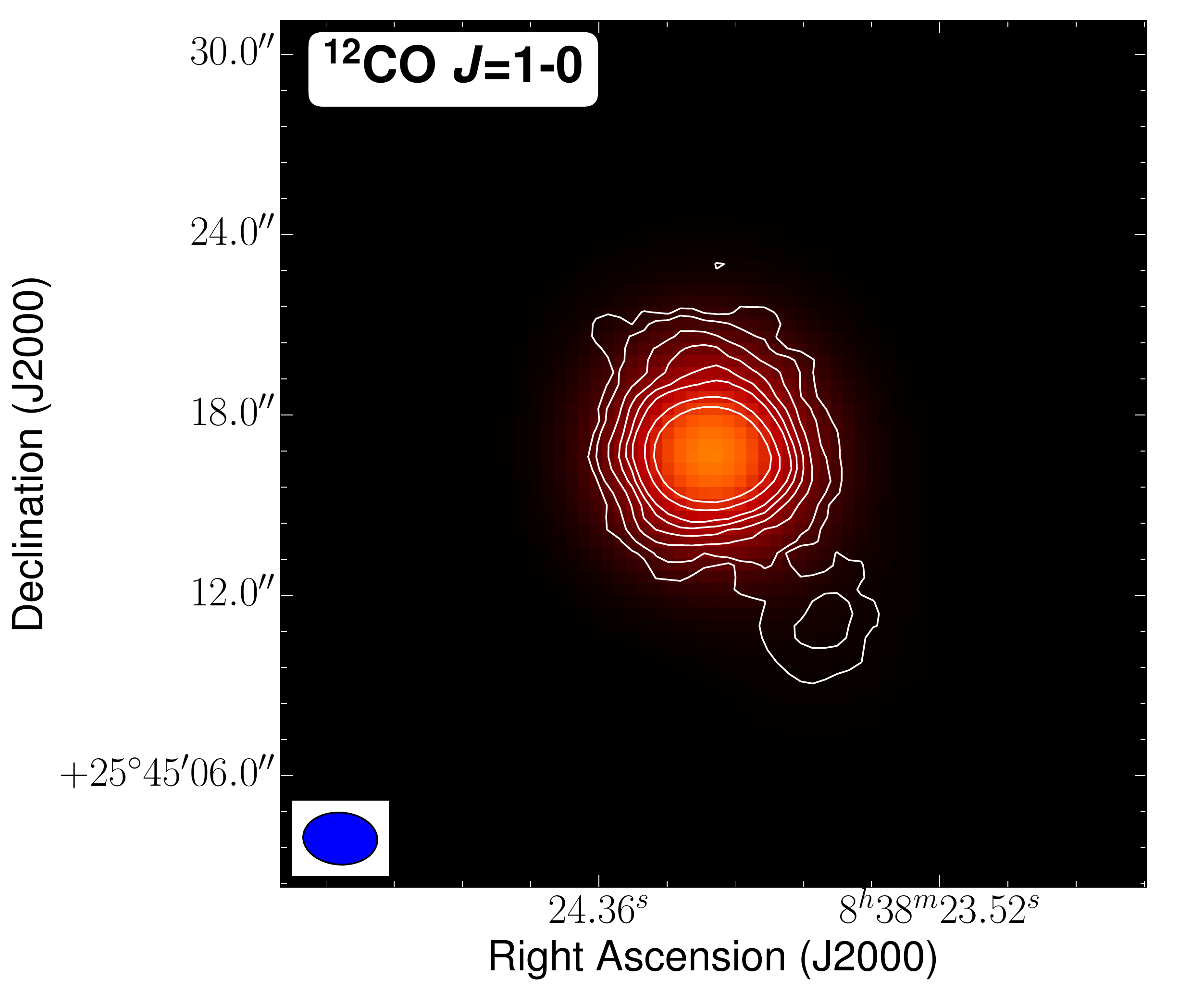}{0.4\textwidth}{(a)}
\fig{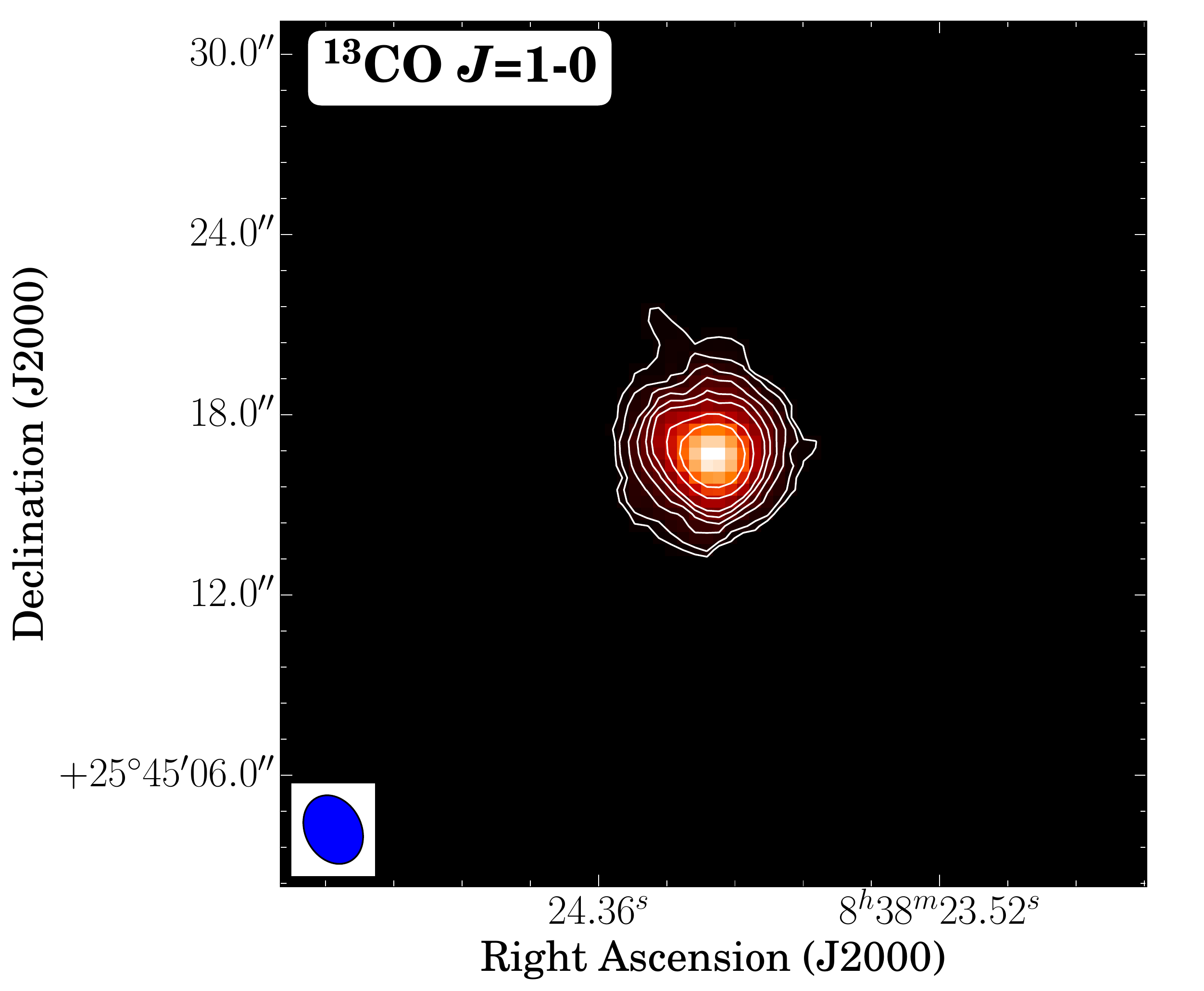}{0.4\textwidth}{(b)}}
\gridline{\fig{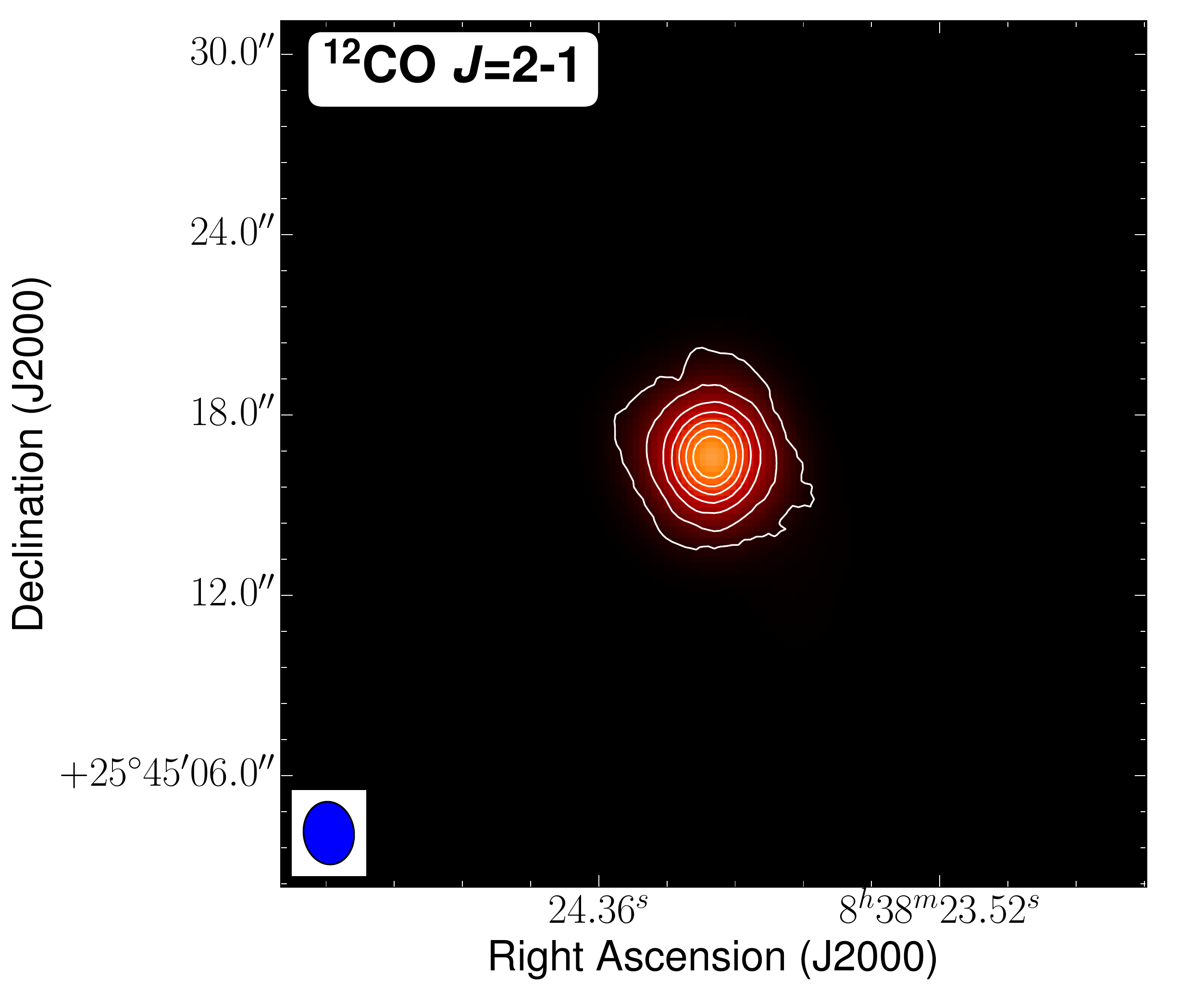}{0.4\textwidth}{(c)}
\fig{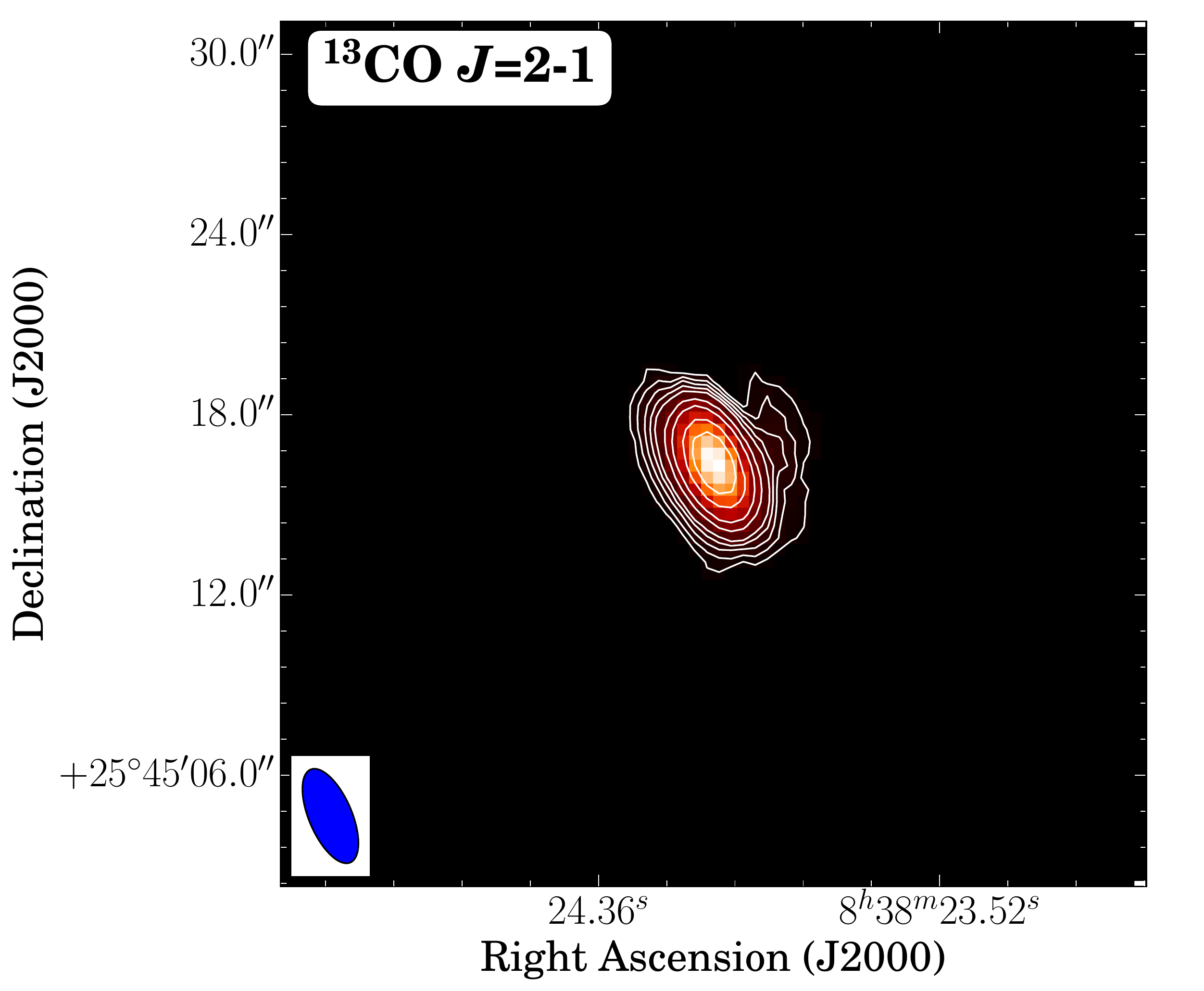}{0.4\textwidth}{(d)}}
\gridline{\fig{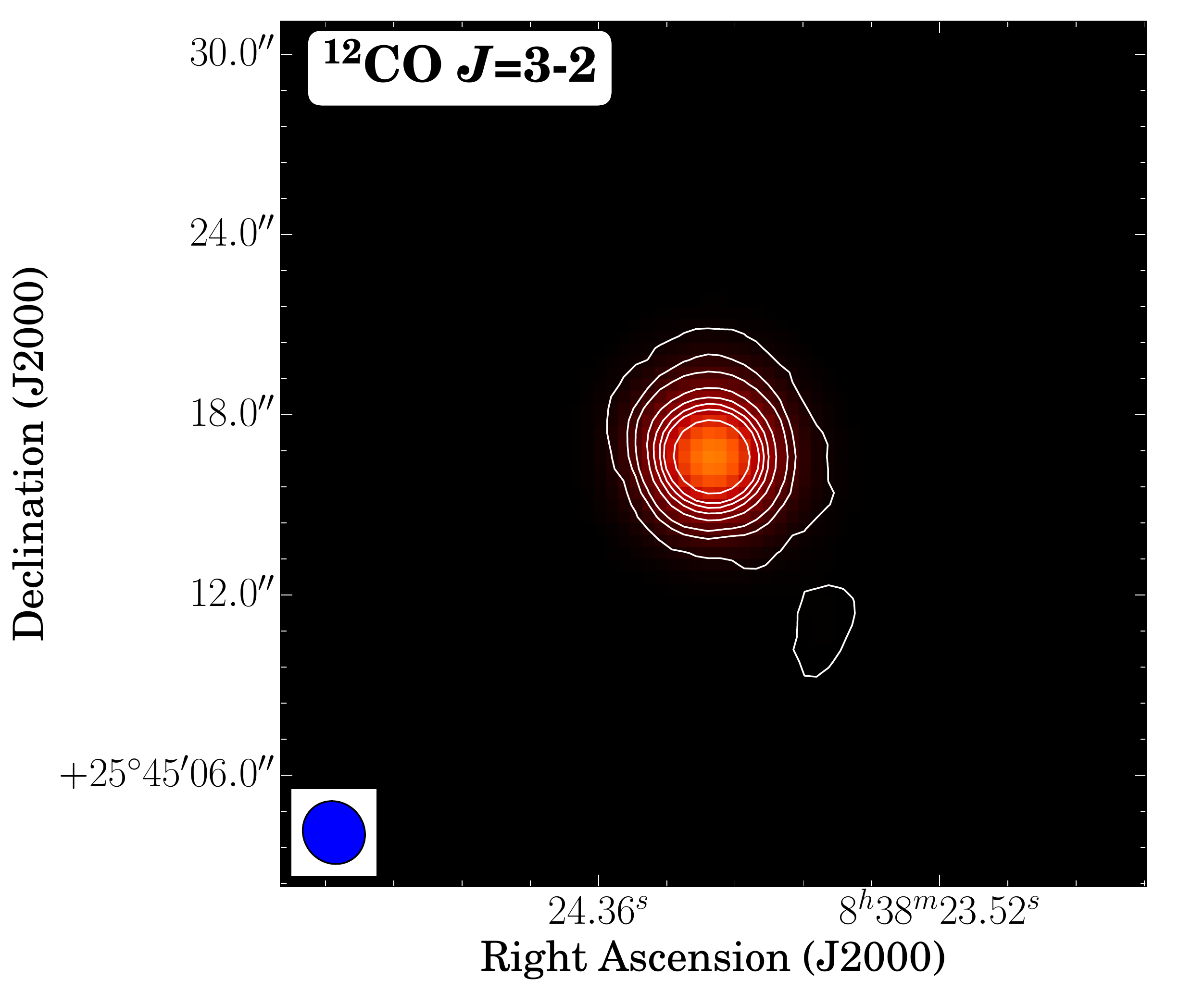}{0.4\textwidth}{(e)}
\fig{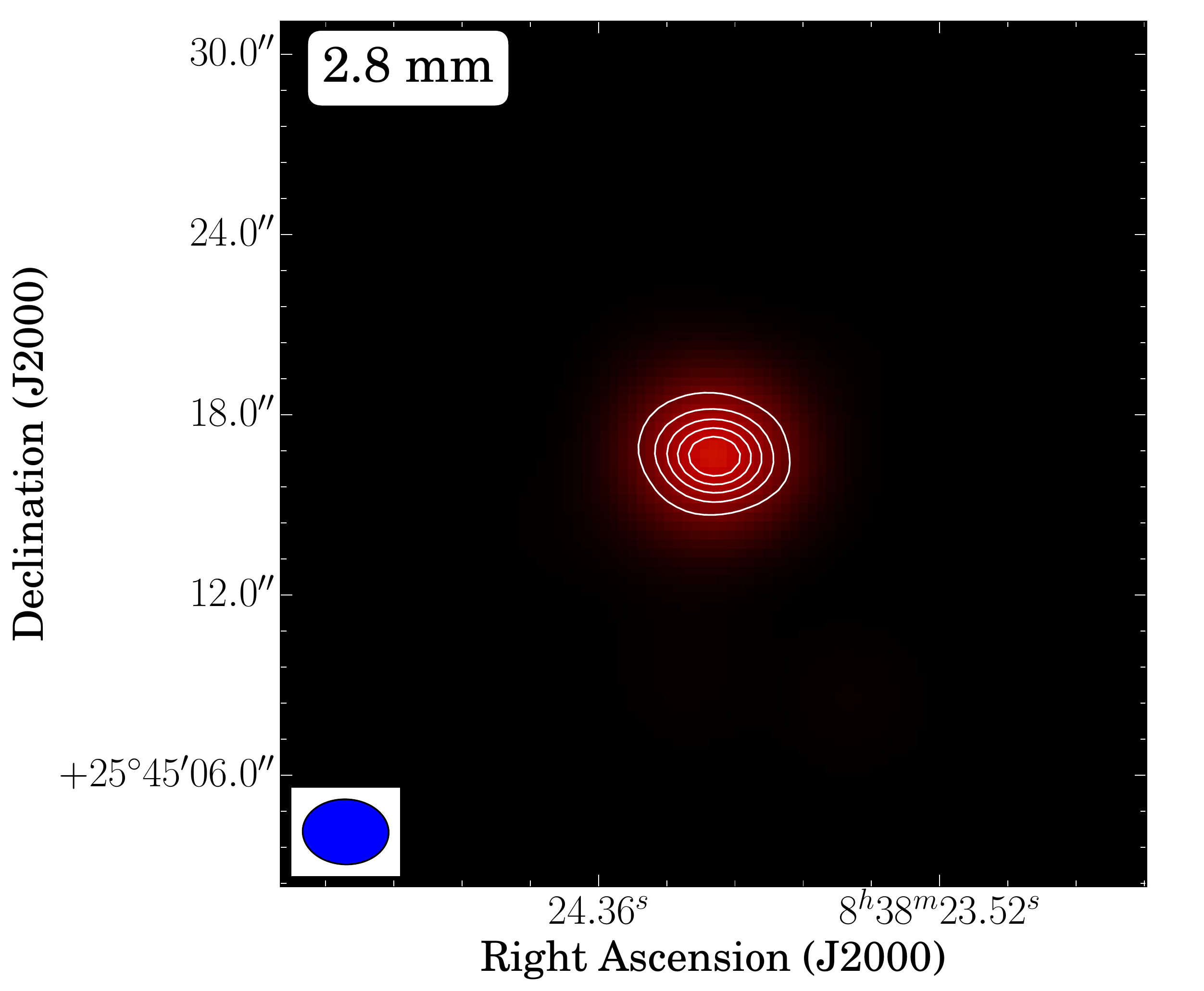}{0.4\textwidth}{(f)}}
\caption{NGC 2623 integrated intensity maps: \textbf{(a)} CARMA \coone. Contours correspond to (1, 2, 4, 6, 
8, 10, 15, 20) $\times$ 2$\sigma$ (= 0.95 Jy beam$^{-1}$ \kms). \textbf{(b)} ALMA \tcoone. Contours 
correspond to [1, 2, 4, 6, 8, 10, 15, 20] $\times$ 1$\sigma$ (= 0.055 Jy beam$^{-1}$ \kms). 
\textbf{(c)} SMA \cotwo\ map. Contours correspond to (1, 5, 10, 20, 25, 30) $\times$ 2$
\sigma$ (= 4.0 Jy beam$^{-1}$ \kms). \textbf{(d)} ALMA \tcotwo. Contours correspond to [2,4, 6, 8,10, 15, 20, 30, 40] $\times$ 1$\sigma$ (0.145 Jy beam$^{-1}$ \kms). 
\textbf{(e)} SMA \cothree. Contours correspond to (1, 6, 10, 20, 
30, 40, 50, 75) $\times$ 2$\sigma$ (= 2.8 Jy beam$^{-1}$ \kms). 
\textbf{(f)} CARMA 2.8~mm continuum. Contours correspond to (5, 10, 15, 20, 25) $\times$ 2$\sigma$ (= 0.46 mJy beam$^{-1}$).
The ellipse in the bottom left corner of each map represents the synthesized beam. All maps 
have been primary beam corrected.   \label{fig:ngc2623maps}}
\end{figure*}

Imaging was conducted in a similar manner to that of the ALMA and CARMA datasets (see \ref{sec:ALMA}).
We improve the maps of Paper I by including clean boxes around CO emission in each 
channel. For \cotwo\ in Arp~55, we image each individual dataset and also a combined 
one; note that Figure 1 displays the combined dataset image.  Note that Paper I did not detect 
\tcotwo\ emission from Arp 55 and NGC 2623. However, the compact configuration 
observation for Arp 55 detected the \tcotwo\ line. The combined dataset of Arp 55 
resolves out the emission of \tcotwo; therefore, we image only the compact configuration 
dataset for \tcotwo\ (Figure \ref{fig:arp55maps}). 

For measurements of the 880~$\mu$m and 1.4~mm continuum, see Paper I.

\subsection{James Clerk Maxwell Telescope (JCMT) and Short Spacings} 
\label{sec:JCMT}
For Arp 55, we used the JCMT to map a 56\arcsec\ $\times$ 56\arcsec\ area in \cotwo\ 
using the RxA3 receiver (PI: K. Sliwa; M12BC14) on 2012 November 26 and 27. A data 
cube spanning -585 to 585 \kms\ was created using the \verb=Starlink= software 
\citep{Currie2008}. Baseline subtraction was performed using line-free channels on 
both sides of the \cotwo\ line. We bin the datacube to a velocity resolution of 30 \kms. 
The intrinsic unit of antenna temperature ($T_{\rm{A}}^{*}$) was converted using a main 
beam efficiency ($\eta_{\rm{mb}}$) of 0.69 to a main beam temperature ($T_{\rm{mb}}$ 
= $T_{\rm{A}}^{*}$/$\eta_{\rm{mb}}$). The unit of $T_{\rm{mb}}$ is converted to Jy using 
a scaling factor of 22.9 Jy K$^{-1}$. We created an integrated intensity map with a 2$
\sigma$ cutoff (0.2 Jy/beam). We measured the total flux of \cotwo\ of Arp 55 to be 410 $
\pm$ 50 Jy \kms. 

The JCMT \cotwo\ flux measured for Arp 55 is consistent with the total flux measured in 
the SMA map (Table \ref{tab:fluxdata}); therefore, we have recovered the flux on all spatial 
scales for Arp 55 in the \cotwo\ map. The largest angular scale (LAS) to which the \cotwo
\ observations are sensitive  is $\sim$12\arc\ (Table \ref{tab:summary}). The 
\coone\ observations are sensitive to even larger scales; therefore, we conclude that those 
observations have also recovered all the flux. The same cannot be said about the \cothree\ 
observations which are sensitive up to $\sim$ 4\arc\ scales. \cite{Papadopoulos2012a} 
measured the \cothree\ flux in Arp 55 in a JCMT spectrum to be 775 $\pm$ 130 Jy \kms; 
therefore, we are missing 70 $\pm$ 20$\%$ of the \cothree\ flux. For the analysis, we 
image all Arp 55 maps with baselines above 30.5 k$\lambda$, the shortest baseline in 
the \cothree\ observations, which will ensure that we are analysing the molecular gas on 
similar scales. 

For NGC 2623, Paper I determined that the \cothree\ SMA map, when compared to the 
JCMT measured flux, recovered all the flux on all scales. The LAS to which the \cothree\ 
observations are sensitive is $\sim$10\arc\ (Table \ref{tab:summary}). Since the 
LAS of the \co, \tcoone\ and \cotwo\ maps are greater than that for the \cothree\ 
observations, we conclude that all the flux is recovered in these maps as well.

\newpage
\begin{deluxetable*}{cccccc} %TABLE 2
%\rotate
%\tablecolumns{6}
\tablewidth{0pt}
%\tabletypesize{\scriptsize}
\tablecaption{Observational Data \label{tab:fluxdata}}
\tablehead{
\colhead{Parameter} & \colhead{Arp 55} & \colhead{Arp 55NE} & \colhead{Arp 55SW} & \colhead{NGC 2623} & \colhead{Units} }
\startdata
\coone\ 			&130 ($\pm$ 3) [$\pm$ 20] & 	80\tablenotemark{a}  ($\pm$ 3) [$\pm$ 20] &	50 ($\pm$ 2) [$\pm$ 8] 	&127 ($\pm$ 6) [$\pm$ 19] & Jy  \kms\	\\	
\cotwo\			&400 ($\pm$ 10) [$\pm$ 80] &285 ($\pm$ 7) [$\pm$ 57] 				&115 ($\pm$ 4) [$\pm$ 20 	&290 ($\pm$ 20) [$\pm$ 58] &Jy  \kms\ \\
\cothree\			& 215 ($\pm$ 20) [$\pm$ 43] &135 ($\pm$ 10) [$\pm$ 27] 			&80 ($\pm$ 9) [$\pm$ 16] 	&650 ($\pm$ 80) [$\pm$ 130] &Jy  \kms\ \\
\tcoone\			&3.8 ($\pm$ 0.3) [$\pm$ 1.1] &1.35 ($\pm$ 0.3) [$\pm$ 0.4] 			&2.45 ($\pm$ 0.3) [$\pm$ 0.8]	&3.1 ($\pm$ 0.3)  [$\pm$ 0.3]&Jy  \kms\ \\
\tcotwo\			&17 ($\pm$ 2) [$\pm$ 5] &13 ($\pm$ 2) [$\pm$ 4]					& 4 ($\pm$ 2) [$\pm$ 1]	&11.2 ($\pm$ 1.5) [$\pm$ 1.7] &Jy  \kms\ \\
$L_{\rm{CO(1-0)}}$	& 9.4 $\pm$ 1.9 &5.8\tablenotemark{a} $\pm$ 1					& 3.6 $\pm$ 0.7			& 2.1 $\pm$ 0.4&10$^{9}$ K \kms\ pc$^{2}$ \\
\mmol\tablenotemark{b}&7.5 $\pm$ 1.5 &4.6\tablenotemark{a} $\pm$ 0.9				& 2.9 $\pm$ 0.6			& 1.6 $\pm$  0.1 &10$^{9}$ \msol \\
\enddata
\tablenotetext{}{--Note: Uncertainties in curved and squared brackets represent measurement and calibration uncertainties, respectively.}
\tablenotetext{a}{Includes the flux from the tail and blob.}
\tablenotetext{b}{\mmol\ = 0.8$L_{\rm{CO(1-0)}}$, assuming \alphaco\ = 0.8 \alphacou\ \citep{Downes1998}}
\end{deluxetable*}

\section{The \co\ Morphology} 
\label{sec:morphology}
\begin{figure*}[htbp] %%%FIGURE 3 HST ARP 55 with SPECTRA of the tail and blob
\centering

\gridline{\fig{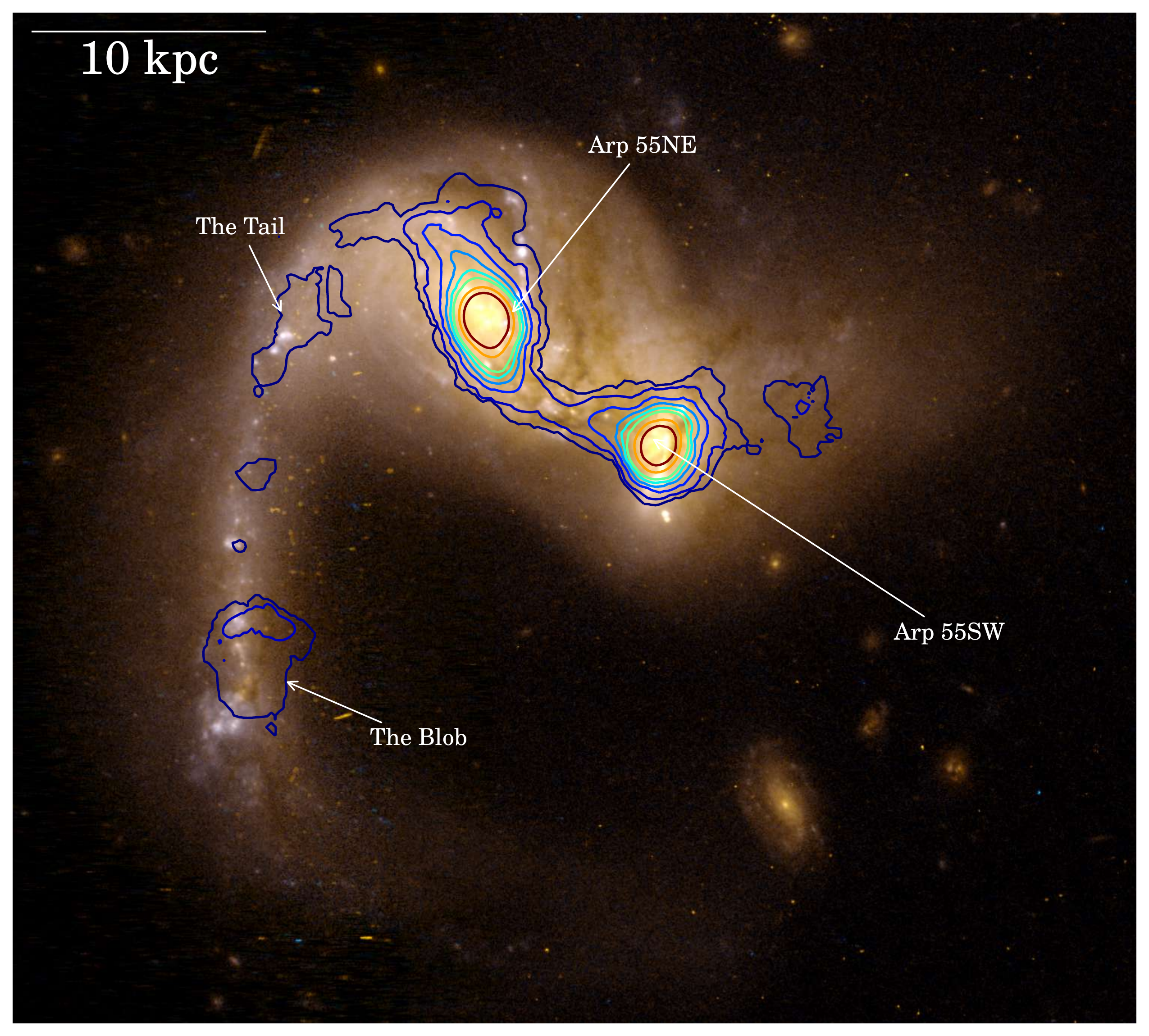}{0.4\textwidth}{(a)}
\fig{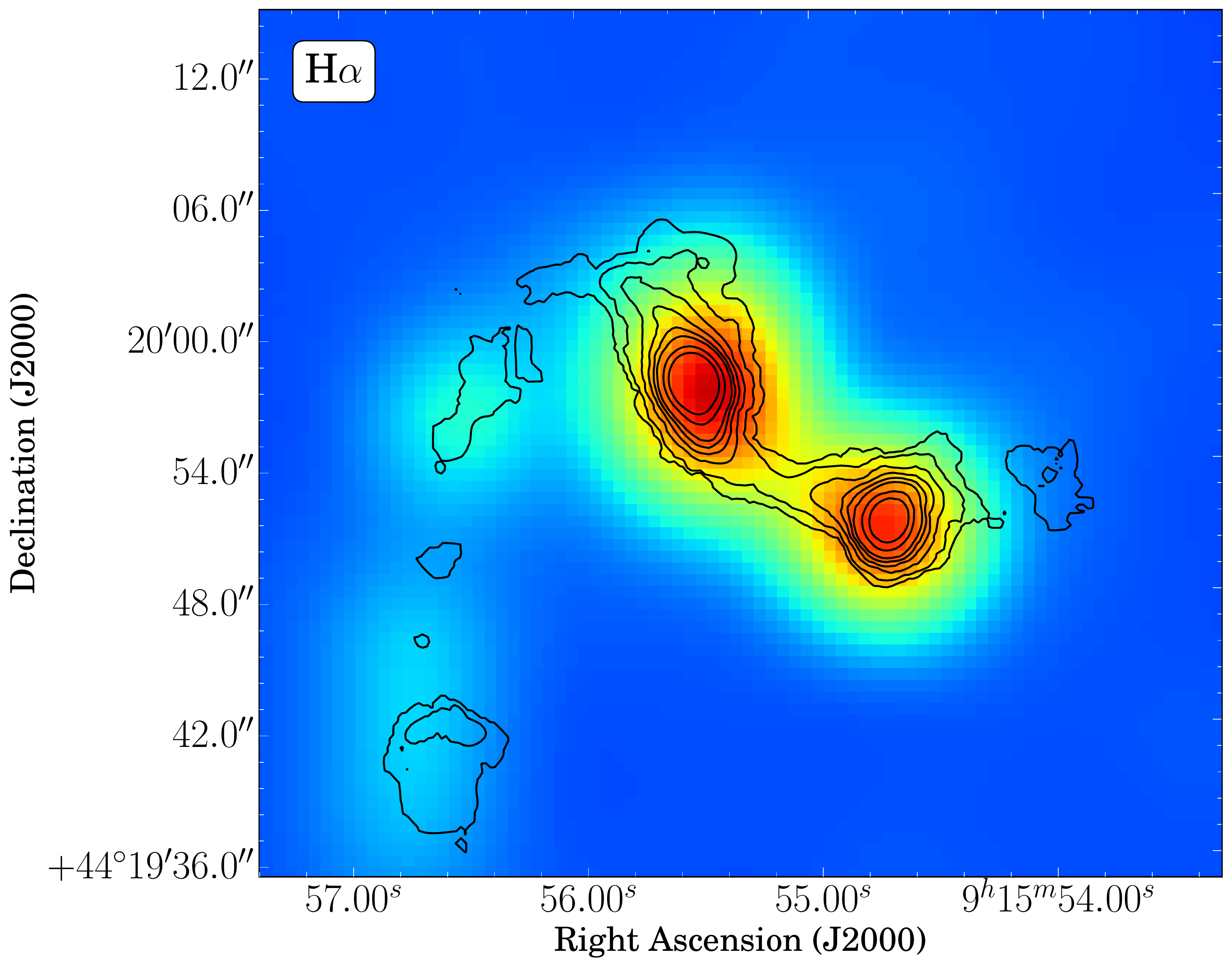}{0.4\textwidth}{(b)}}

\gridline{\fig{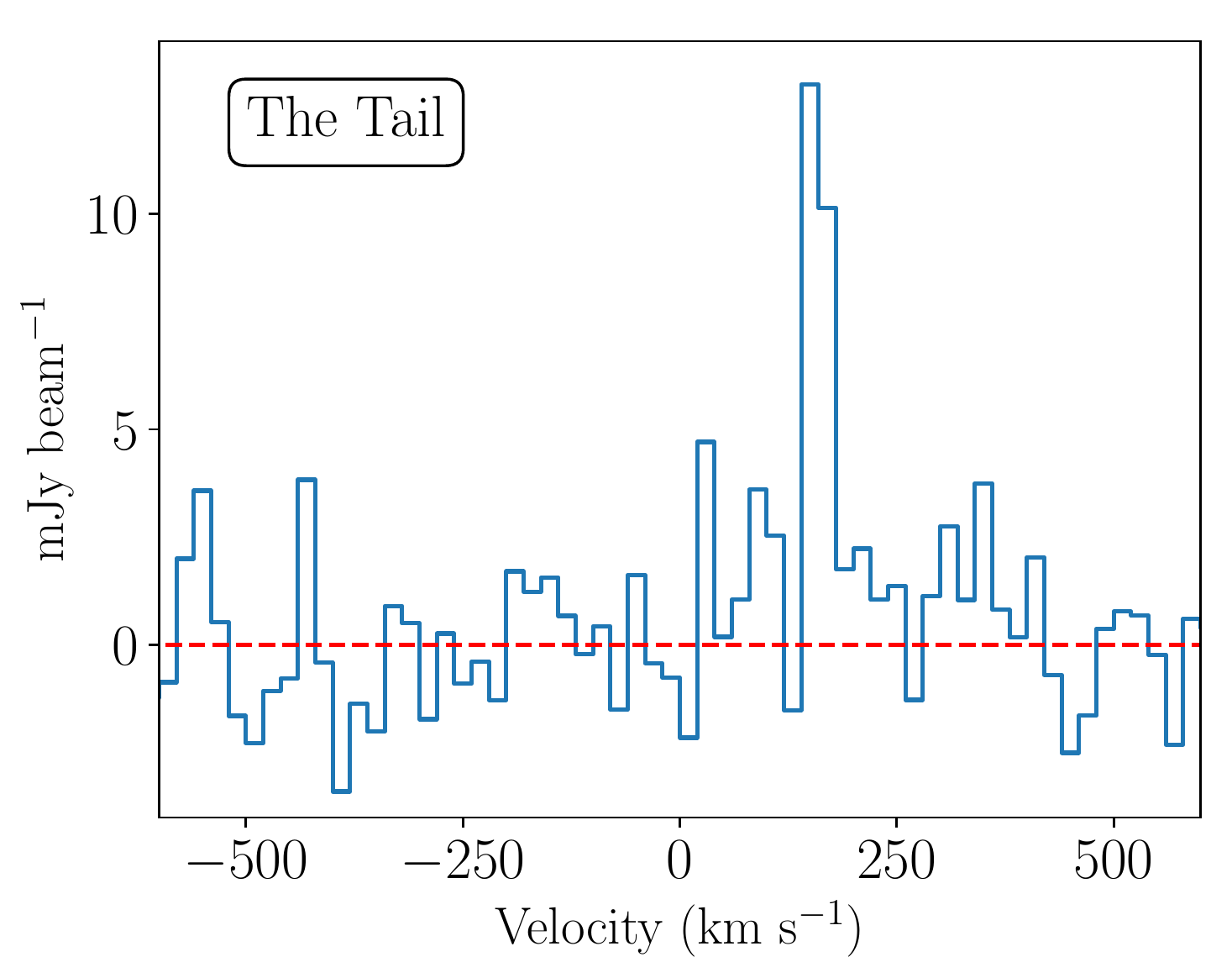}{0.4\textwidth}{(c)}
\fig{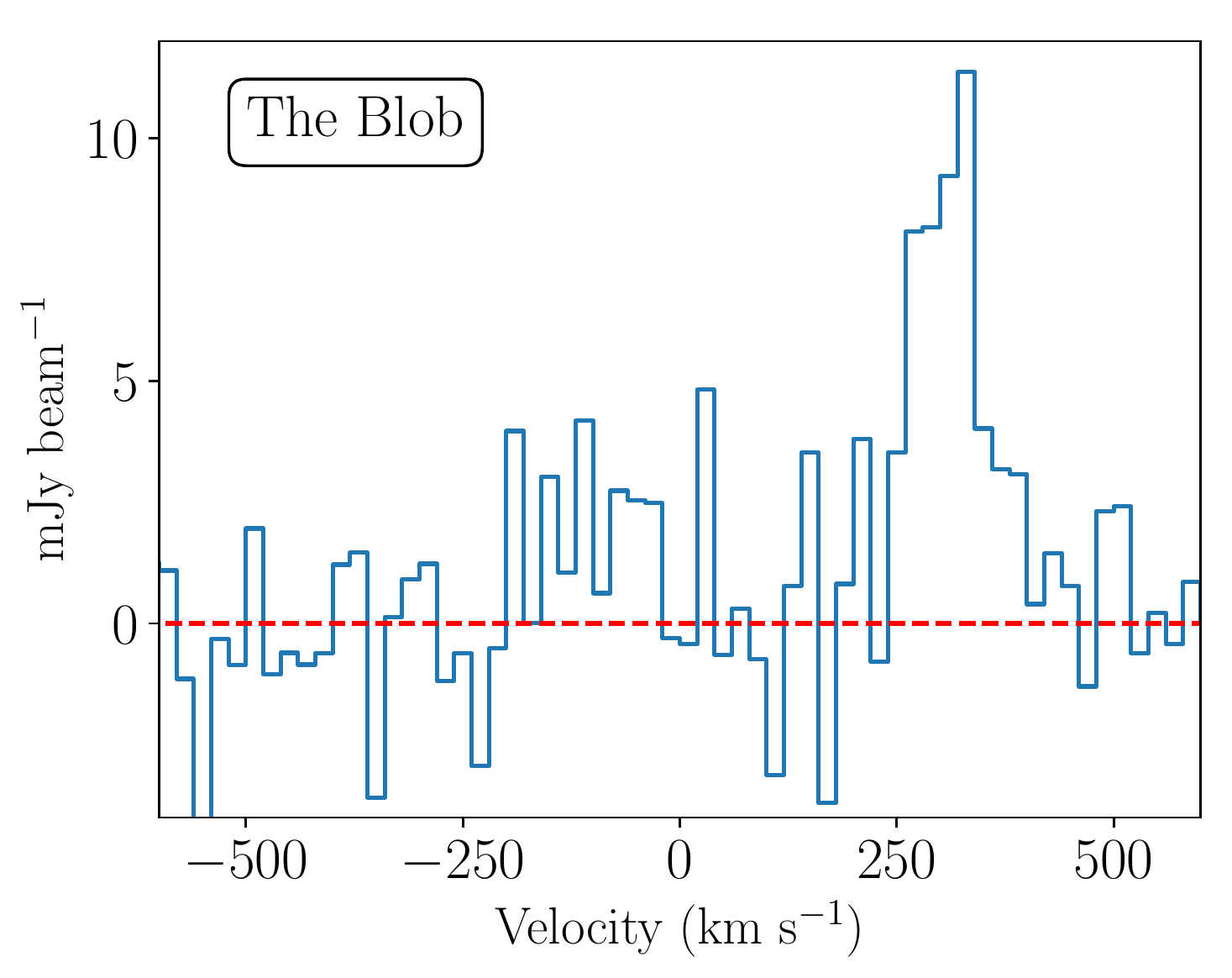}{0.4\textwidth}{(d)}}

\caption{ (a) \textit{Hubble Space Telescope} (HST) colour image with \coone\ contours overlaid.. (b) H$\alpha$ map 
(Hattori et al. 2004) of Arp 55 with CARMA \coone\ contours overlaid. (c) Average CO line 
profiles of the tail and (d) the blob averaged over a 4\arcsec\ apertured centered at ($\alpha_{\rm{J2000}}$ = 09$^{\rm{h}}$15$^{\rm{m}}$56$^{\rm{s}}$40, $\delta_{\rm{J2000}}$ = +44$^{\circ}$19$^{\prime}$58\arcsec.97) and ($\alpha_{\rm{J2000}}$ = 09$^{\rm{h}}$15$^{\rm{m}}$65$^{\rm{s}}$40, $\delta_{\rm{J2000}}$ = +44$^{\circ}$19$^{\prime}$41\arcsec.29), respectively. \label{fig:arp55hst} }
\end{figure*}

Arp 55 was first observed interferometrically by \cite{Sanders1988}  resulting in a \coone\ map 
with a resolution of 9\arc\ $\times$ 7\arc\ and a total flux of 83 Jy \kms. Our CARMA map has 
better angular resolution, greater sensitivity and higher total flux recovered. In addition to 
detecting the two nuclei, we also detect a tail-like feature in the north-eastern galaxy and a 
blob in the tidal tail (Figure \ref{fig:arp55hst}). Both regions have active star formation as seen 
in the H$\alpha$ map of \cite{Hattori2004} (Figure \ref{fig:arp55hst}). The tail and blob have 
not been detected in \cotwo, but we believe this to be an issue of sensitivity and not missing 
flux. Assuming a line ratio of \co\ (2-1)/(1-0) = 1.5 (K units; see Appendix \ref{sec:lineratio}) for the blob, with the sensitivity of the \cotwo\ map 
and taking into account the drop off in sensitivity towards the edges of the primary beam, the 
peak of the blob would be at the $\sim$2$\sigma$ level. 

The blob has a total flux of $\sim$ 9 Jy \kms\ which translates to a total molecular gas mass of $\sim$ 5.2 ($\frac{\alpha_{\rm{CO}}}{0.8})$ $\times$ 10$^{8}$ \msol\ assuming a conversion factor of 0.8 \alphacou\ \citep{Downes1998}. While it is unclear whether a U/LIRG-like conversion factor is appropriate for the tidal tail blob, we note that Galactic conversion factor would give a total molecular gas mass $\sim$ 3 $\times$ 10$^{9}$ \msol. 
 
This is an interesting region as it may lead to a tidal dwarf galaxy, a type of small galaxy that 
forms from material ejected from the disks of colliding spiral galaxies \citep[e.g.][]{Braine2001, 
Braine2000}. They are usually found at the ends of long tidal tails, as is the case with Arp 55, 
and the molecular gas emission peaks near the peak of atomic hydrogen (HI) emission 
\citep{Braine2000} suggesting that the molecular gas is formed from the HI. \cite{Thomas2002} 
observed Arp 55 in HI using the VLA; however, there was no detection over the entire source with a 3$\sigma$ noise-level 
upper limit to the total HI mass of 8.9 $\times$ 10$^{9}$ \msol\ assuming a velocity line width of 350 \kms. A potential tidal dwarf galaxy 
has also been found in the LIRG VV 114 \citep{Saito2015}. 

\begin{figure*}[htbp] %%%FIGURE 3 HST NGC 2623 with SPECTRA of the blob
\centering

\gridline{\fig{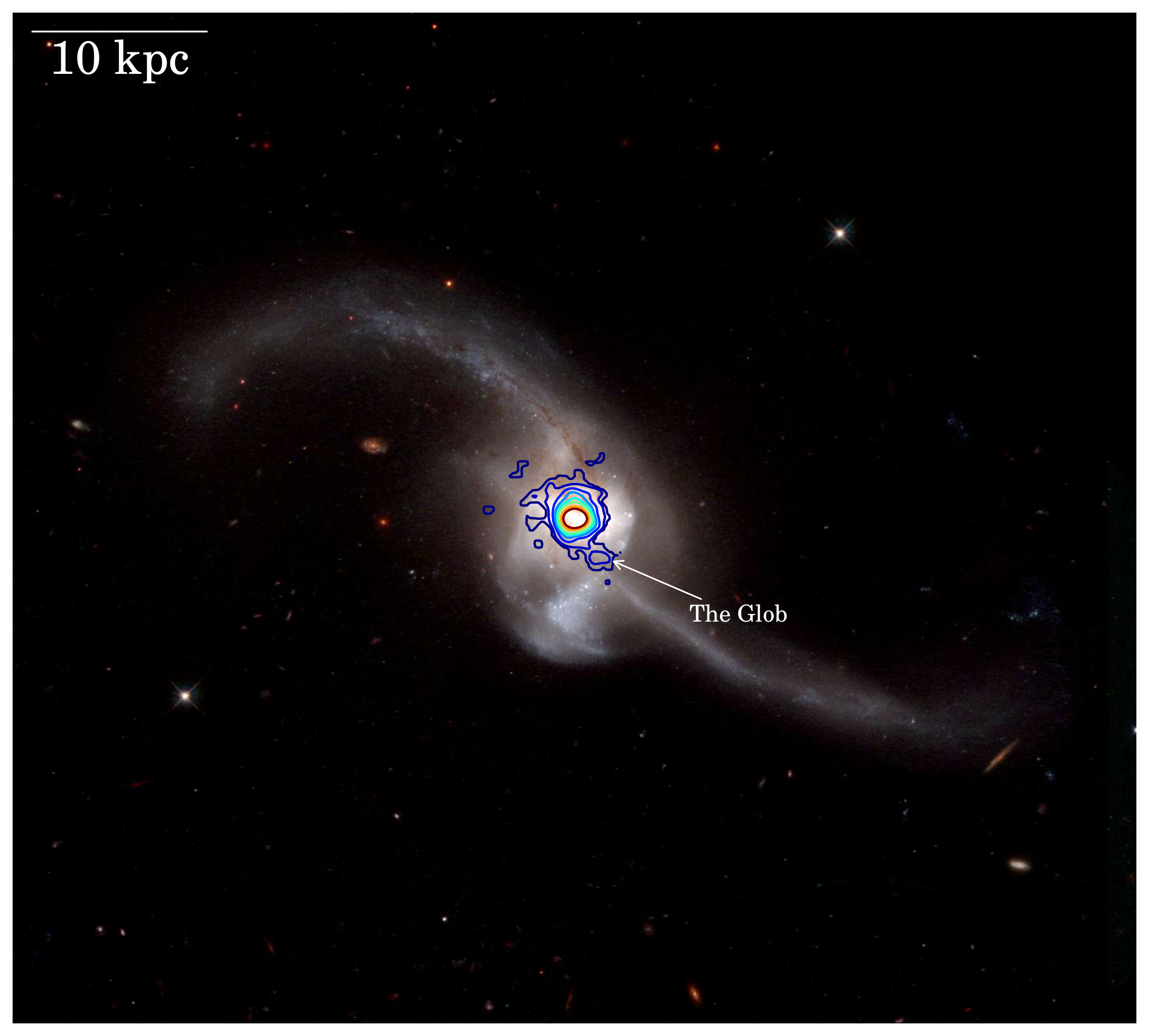}{0.4\textwidth}{(a)}
\fig{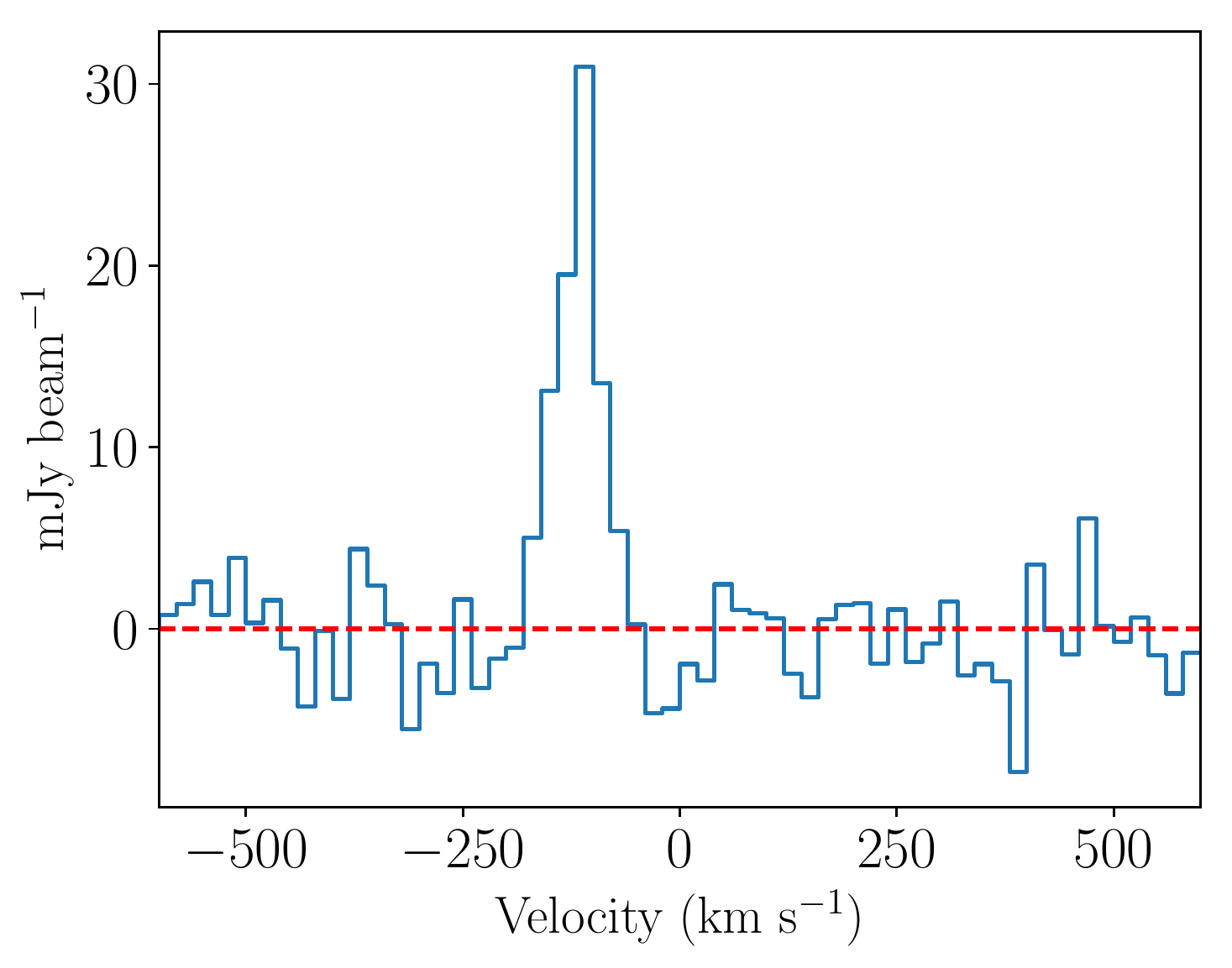}{0.4\textwidth}{(b)}}

\caption{(a) HST color image of NGC 2623 with CARMA \coone\ contours overlaid. 
(b)  The \coone\ line profile of the Glob southwest of the central compact nucleus, averaged over a 4\arcsec\ $\times$ 3\arcsec\ aperture centred at ($\alpha_{\rm{J2000}}$= 08$^{\rm{h}}$38$^{\rm{m}}$23$^{\rm{s}}$.788, $\delta_{\rm{J2000}}$ = +25$^{\circ}$45$^{\prime}$10\arcsec.782).\label{fig:ngc2623hst}}
\end{figure*}

NGC 2623 was observed in \coone\ by \cite{Bryant1999} resulting in a map with a 
resolution of 3.5\arc\ $\times$ 2.4\arc\ with a total flux of 153 Jy \kms. The flux in our CARMA 
map is lower than that of \cite{Bryant1999}; however, we attribute this difference to the 
calibration uncertainty. The \coone\ emission is compact, similar to that of \cite{Bryant1999}; 
however, we do detect an off-nuclear concentration of molecular gas (the Glob; Figure \ref{fig:ngc2623hst}) to the south-west that is not seen in the map of \cite{Bryant1999}. We also detect the Glob in the \cothree\ map, confirming that it is a real feature. The Glob has a flux of 5.2 Jy \kms\ corresponding to a molecular gas mass of 6.9 ($\frac{\alpha_{\rm{CO}}}{0.8}$) $\times$ 10$^{7}$ \msol.  The velocity at the peak position of the Glob is $\sim$160~\kms; using the diameter and dynamical mass of NGC 2623 measured from the CO emission (see Section \ref{sec:alphaco} and Table \ref{tab:linewidth}), the escape velocity is estimated to be $\sim$ 400~\kms. This escape velocity indicates that the Glob will not escape the system.

%%%%%%%%%%%%%%%%%%%%%%%%%%%%%%%%%%%%%%%%%%%%%%%
%%%%%

%%%%%%%%%%%%

\section{Molecular Gas Properties} \label{sec:moleculargas}
\subsection{Radiative Transfer Modeling}\label{sec:RADEX}

RADEX is a commonly used code that can, as stated in \cite{vanderTak2007}, ``calculate the intensities of atomic and molecular lines produced in a uniform medium, based on statistical equilibrium calculations involving collisional and radiative processes and including radiation from background sources. Optical depth effects are treated with an escape probability method." The code can be used to generate model points with the following parameters:
kinetic temperature (\tkin), column density of a molecular species X per unit line width (N$_{X}$/$\Delta$V) and volume density of molecular hydrogen (\nhtwo). In combination with a Bayesian likelihood code, we can constrain the most probable set of physical conditions of the molecular gas. We use the code published in \cite{Kamenetzky2014}\footnote{https://github.com/jrka/pyradexnest} that implements a nested sampling algorithm that focuses on high likelihood regions efficiently constraining the parameters. In addition to the three parameters mentioned above, the flux is permitted to scale uniformly down by an area filling factor \ff ($\leq$ 1). Please see \cite{Kamenetzky2014} for more details. 

To further aid the model fits and prevent unphysical conditions, we implement three priors:
\begin{enumerate}
\item The dynamical mass is used as an upper limit to the total mass that can be contained within the beam. This prior helps constrain the column density solution. The dynamical mass is measured using the source diameter deconvolved from the synthesized beam, the full-width half-maximum (FWHM) of the \co\ line profile and assuming a radially decreasing density profile \citep[][see Table \ref{tab:linewidth}]{Wilson1990}. 
\item The column length is used to constrain the column and volume densities at the higher and lower end, respectively. We estimate the column length from the deconvolved source diameter, assuming a spherical geometry. While a spherical geometry is likely not correct, this still offers an upper limit to the column length. For Arp 55, since the source does not appear to be edge-on and therefore, the column length is more reminiscent of the disk scale height, we take half of the deconvolved source size because it is very unlikely that the column length is 1.5~kpc. 
\item The final prior limit the optical depth, $\tau$, within the range of [0,100]. The lower limit is due to the fact that carbon monoxide is most likely not a maser ($\tau$ $ <$ 0) and the upper limit is recommended by the RADEX documentation. 
\end{enumerate}

%\newpage
\begin{deluxetable*}{cccccc}%%%TABLE 2
%\rotate
%\tablecolumns{6}
\tablewidth{0pt}
\tablecaption{Source Size and Dynamical Mass \label{tab:linewidth}}
\tablehead{\colhead{Galaxy} & \colhead{Velocity FWHM}&  \multicolumn{2}{c}{Deconvolved 
Source Diameter} & \colhead{$M_{\rm{dyn}}$\tablenotemark{a}} & \colhead{\alphaco$_{\rm{, 
dyn}}$\tablenotemark{b}} \\
\colhead{} & \colhead{(km s$^{-1}$)}&  \colhead{(arcsec)} & \colhead{(kpc)} & 
\colhead{(10$^{9}$ $M_{\odot}$)} & \colhead{(\alphacou)} }
\decimalcolnumbers
\startdata
Arp 55NE&	220 $\pm$ 20	&	(2.6 $\pm$ 0.2) $\times$ (1.2$\pm$ 0.2)	&	(2.2 $\pm
$ 0.2) $\times$ (1.0 $\pm$ 0.2)	& 	7.2 $\pm$ 1.7 & 1.4 $\pm$ 0.4	\\
Arp 55SW	&	260 $\pm$ 20	&	(2.0 $\pm$ 0.2) $\times$ (1.3 $\pm$ 0.2)	&	(1.7 $\pm
$ 0.2) $\times$ (1.1 $\pm$ 0.2)	&	9.2 $\pm$ 2.1 & 2.6 $\pm$ 0.8	\\
NGC 2623&	290 $\pm$ 20	&	$\leq$1.4 $\pm$ 0.1	&	$\leq$0.56 $\pm$ 0.04 &	$\leq
$4.7 $\pm$ 0.7 & $\leq$2.2 $\pm$ 0.5	\\
\enddata
\tablenotetext{a}{ $M_{\rm{dyn}}$ = 99$\Delta$V$^{2}$D(pc) \citep{Wilson1990}
where $\Delta$V is the velocity FWHM and D is the diameter of the region in parsecs.}
\tablenotetext{b}{\alphaco$_{\rm{, dyn}}$ = $M_{\rm{dyn}}$/$L_{\rm{CO(1-0)}}$. 
$L_{\rm{CO(1-0)}}$ is taken from Table \ref{tab:fluxdata}. For Arp 55NE, we subtract out the 
luminosity corresponding to the blob. }
\end{deluxetable*}

\subsubsection{Arp 55}\label{sec:arp55model}

We model all the emission lines (Table \ref{tab:fluxdata}) from Arp~55 at the \cothree\ peak intensity position of the two nuclei at the resolution of the \coone\ map ($\sim$ 1.4~kpc). We smooth the angular resolution of the observations using a Gaussian taper during imaging. The native resolution of the \tcotwo\ map is coarser than the map of \coone; therefore, we estimate the flux of \tcotwo\ at the resolution of \coone\ using an average \tco\ (2-1/1-0) line ratio of 1.6 (K units; see Appendix \ref{sec:lineratio}). Using the \tco\ line ratio will introduce fewer uncertainties because we assume the two \tco\ transitions have similar filling factors. We also implement a 30$\%$ uncertainty for \tco\ which covers the observed range of \tco\ (2-1/1-0) within the regions with $\geq$3$\sigma$ detections. Using the \tco\ line ratio is appropriate as both \tco\ lines likely originate from the same region with similar sizes. The emission from both Arp 55 nuclei is consistent with cold ($\sim$ 10-20~K), dense ($>$ 10$^{3})$ cm$^{-3}$) molecular gas (Table \ref{tab:results}, Figure \ref{fig:radexsolution}). The fit is similar to another relatively younger merger, NGC 4038/9 \citep{Schirm2014}. Note we do not model the tidal tail and blob due to the lack of detections of the other emission lines (i.e. \tco\ and \cotwo\ and $J$=3-2).

\subsubsection{NGC 2623}\label{sec:n2623model}
\begin{figure*}[htbp] %%%FIGURE 5 NGC 2623 line ratio maps
\centering
\includegraphics[scale=0.8]{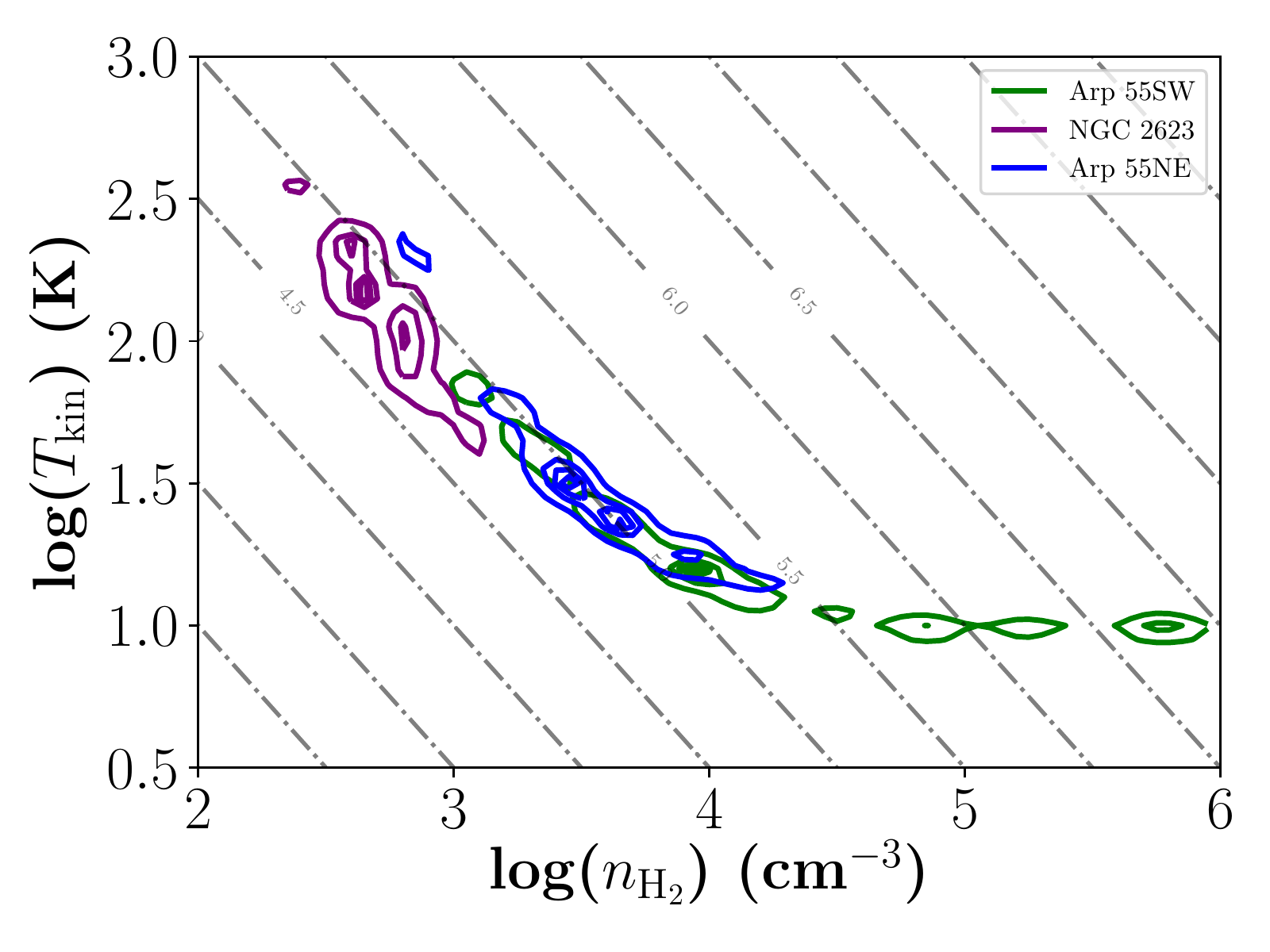}
\caption{Probability distributions for the temperature (\tkin) versus volume density (\nhtwo) with colored contours corresponding to 55, 80, 90 and 95\% of the most likely solution. Dash-dot lines represent log(Pressure) [cm$^{-3}$ K]. \label{fig:radexsolution}}
\end{figure*}

The emission lines from NGC 2623 were modeled at the \cothree\ peak position at a compromise resolution of 2.5\arcsec\ ($\sim$~1 kpc). As with Arp~55, the \tcotwo\ map resolution is too coarse and we estimate the flux at 2.5\arcsec\ using an average \tco\ (2-1/1-0) line ratio of 0.85 (K units; see Appendix \ref{sec:lineratio}). The emission is consistent with warm ($\sim$ 110 K), moderately dense ($<$ 10$^{3.0}$ cm$^{-3}$) molecular gas (Table \ref{tab:results}, Figure \ref{fig:radexsolution}). This best fit has similarities to the fit for Arp 220 on global scales \citep{Rangwala2011}; however, the best fit temperature is twice that for Arp 220. 

\section{Discussion}\label{sec:discussion}
\subsection{\xco\ Abundance Ratio} \label{sec:abundance}
Analyses of several local LIRGs and even high redshift starbursts have shown that their \xco\ abundance ratios are unusually high \citep[$\geq$90;][]{Henkel2010,Henkel2014,Sliwa2013,Sliwa2014,Papadopoulos2014,Tunnard2015b} when compared to the Galaxy. In the Galaxy, \xco\ varies radially ranging from 30 near the center to $>$100 at larger radii \citep[e.g.][]
{Langer1990,Milam2005}. During the merger process, gas is driven towards the 
nuclear regions and if pristine high \xco\ valued gas is funnelled into the centers, \xco\ can be 
enhanced in the centers by a factor of $\sim$2 \citep[e.g.][]{Casoli1992,Rupke2008}. This mechanism, if dominant, suggests that the \xco\ value cannot significantly exceed 100. Another mechanism to increase \xco\ is via the 
enrichment of $^{12}$C through stellar nucleosynthesis. The $^{12}$C atom is produced in intermediate and high-mass stars via the 3$\alpha$ process that can (for example, with an 8~\msol\ star) start to enrich the ISM in $\sim$ 10$^{7}$ years. The $^{13}$C atom, 
on the the other hand, is an intermediary product during the CNO cycle that is transformed into $^{14}$N; however, a significant amount is lifted to the envelopes of low/intermediate-mass stars during the red giant phase as part of a dredge-up process \citep{Wilson1992}. This process will start to affect the $^{13}$C abundances in the ISM in $\sim$10$^{9}$ years. The chemical models of \cite{Vigroux1976} show that the low mass stars start to significantly influence the ISM in 10$^{9}$ years. 

NGC 2623 is consistent with this trend with an unusually high \xco\ value ($\sim$ 250; Table \ref{tab:results}).  The inflow of pristine gas cannot fully explain this abundance suggesting another mechanism is required such as the stellar nucleosynthesis enrichment. Further testing can be implemented using other isotopologues such as oxygen, where $^{17}$O is believed to be produced in low/intermediate mass stars \citep[e.g.][]{Sage1991,Wilson1992} and should be low in abundance while $^{18}$O is believed to be mainly produced in massive stars and should be increased in abundance. The simple modeling of \cite{Henkel1993} show that the abundance ratio of [$^{16}$O]/[$^{18}$O] significantly decreases around 5-10 $\times$ 10$^{7}$ Myr after the merger starts. Arp~220 shows an enhanced brightness temperature line ratio of $^{13}$CO/C$^{18}$O $\sim$ 1 \citep{Greve2009,Matsushita2009}, suggesting an increase in C$^{18}$O abundance due to stellar nucleosynthesis. We note that normal disk galaxies show $^{13}$CO/C$^{18}$O $\sim$ 6 \citep{JD2017}.

Interestingly, Arp 55 does not conform to the trend, with the most probable \xco\ value of 15 -- 30, similar to the value at the center of the Galaxy \citep{Langer1990}.  Arp 55 is at a relatively earlier merger stage where gas inflow is likely still occurring. The \coone\ map shows extended emission (Figure \ref{fig:arp55maps}) and it may be that not enough of the outer radii high \xco\ valued gas has reached the inner kiloparsec region to drive up the abundance ratio. If such a scenario is plausible, we should see a radial gradient in the abundance ratio; however, higher resolution and more sensitive \tco\ observations are required for such an analysis.

 Another plausible scenario is fractionation. Carbon isotope exchange is possible via the reaction
\begin{equation}\label{eqn:frac}
^{13}\rm{C^{+}} + \rm{^{12}CO} \rightleftharpoons \rm{^{12}C^{+}} + \rm{^{13}CO} + 35\rm{K}
\end{equation}
as predicted by \cite{Watson1976}. The forward reaction dominates in cold environments favoring the formation of \tco. In hot environments, both the forward and reverse reactions are about equally probable \citep[e.g.][]{Roueff2015} and will not affect the relative abundance ratio significantly. The RADEX solution suggests that the molecular gas is cold which will favor the conversion of \co\ to \tco\ in Arp 55. If this is the dominant mechanism, the \xco\ abundance ratio in Arp 55 is not a good tracer of the carbon isotope ratio, [$^{12}$C]/[$^{13}$C] in this system. To test this scenario, other carbon bearing molecular species observations are need such as cyanide \citep[CN;][]{Milam2005,Henkel2014}.

\subsection{CO-to-H$_{2}$ Conversion Factor: \alphaco} \label{sec:alphaco}

The \alphaco\ factor is important to determine the amount of fuel for star formation in galaxies. 
Since the first high-resolution observations of LIRGs \citep[e.g.][]{Downes1993}, adopting the 
Galactic \alphaco\ value (4.3 \alphacou; see \citealt{Bolatto2013} for a review)
\nocite{Bolatto2013} resulted in molecular gas masses that were equal to or greater than the 
dynamical masses. \cite{Downes1998} found via radiative transfer models that \alphaco\ in 
several LIRGs is on average a factor of $\sim$5 lower than the Galactic value, with \alphaco\ values ranging from 0.3 to 1.0 \alphacou. If the merging 
galaxies start with a Galactic-like \alphaco, at some point in the merger process the value will 
transition into the LIRG value and some young mergers should show intermediate \alphaco\ value (i.e. between $\sim$ 1 and 4). 

By comparing the \co\ column density (Table \ref{tab:results}) with the peak \coone\ luminosity we 
measure \alphaco\ for Arp 55NE, Arp 55SW and NGC 2623 to be 0.1, 0.08 and 0.7 (3 $
\times$ 10$^{-4}$/\xh) \alphacou, respectively.  Surprisingly, Arp~55 shows an enhanced \alphaco\ even though it is at an early merger stage; however, the exact merger stage is unknown. It is most likely well after first passage and may be coming in for the second passage. This low \alphaco\ value for Arp~55 suggests either that the search for intermediate \alphaco\ values must take place in younger mergers near first passage or the transition to LIRG-\alphaco\ values is rapid.

The largest uncertainty in these values arises from the assumption of the \co\ abundance relative to H$_{2}$, \xh\ 
= [\co]/[H$_{2}$], where we adopt the value of 3 $\times$ 10$^{-4}$, a value found in warm 
star forming clouds \citep[e.g][]{Lacy1994}. The \xh\ value will likely vary from 
galaxy to galaxy with varying metallicities with literature values ranging from 0.8 - 5 $\times$ 10$^{-4}$ \citep[e.g.][]{Watson1985,Frerking1989,Black1990,Lacy1994}. We note that \cite{Downes1998} assume an \xh\ value of 8.5 $\times$ 10$^{-5}$. 
Future observations of several \co\ and \tco\ transitions in giant molecular clouds in 
different environments are needed to perform a line intensity analysis to measure \xh\ 
directly.

We also measure the \alphaco\ value using the dynamical mass (Table \ref{tab:linewidth}), \alphaco
$_{\rm{, dyn}}$. Since the dynamical mass includes the mass of all matter (i.e. stars, gas, dust, 
dark matter, etc.), \alphaco$_{\rm{, dyn}}$ is strictly an upper limit to the true \alphaco\ value:
\begin{equation}
\alpha_{\rm{CO, dyn}} \ge \alpha_{\rm{CO, true}} \propto \frac{N_{\rm{^{12}CO}}}{[\rm{^{12}CO}]/[\rm{H_{2}}]}
\end{equation}
The upper limit is still 1.5-3 times lower than the Galactic value.  By combining $M_{\rm{dyn}}$ within the analysis resolution
and \nco\, we  can also place a lower limit on \xh\ of $\sim$ 1.2 and 2 $\times$ 10$^{-5}$ for Arp 55 
and NGC 2623, respectively. Our assumption of 3 $\times$ 10$^{-4}$ is more than an order of 
magnitude higher than the lower limit. 

\subsection{Molecular Gas Properties as a Function of Merger Stage}\label{sec:mergerstage}

\begin{figure*}[htbp] %%%\ref{mergertempden}
\centering
\includegraphics[ scale=1.1]{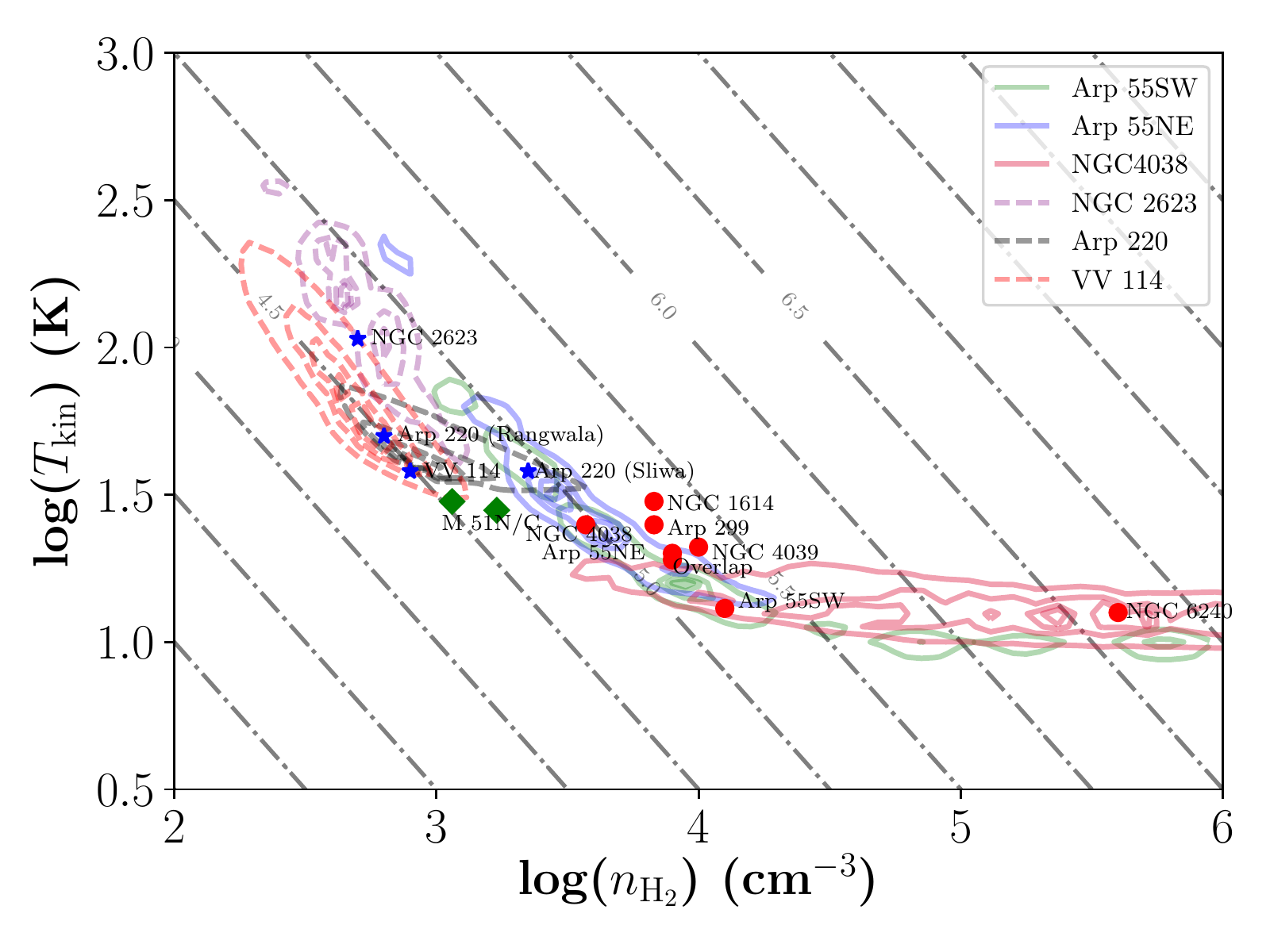}  
\caption{Most probable temperature versus volume densities for several luminous mergers. Data 
points include Arp 220 \citep[][Sliwa $\&$ Downes 2016 submitted.]{Rangwala2011}, Arp 299 \citep{Sliwa2012}, VV 114 \citep{Sliwa2013}, NGC 1614 \citep{Sliwa2014}, NGC 4038/39 + overlap \citep{Schirm2014}, NGC 6240 \citep{Tunnard2015b}, M 51 nucleus and center  (green diamonds; Schirm et al. 2016 submitted), NGC 2623 and Arp 55. Blue stars denote advanced mergers and red circles denote the early/intermediate mergers. Colored contours represent the 55, 80, 90, 95$\%$ most probable solution distribution for Arp 55, NGC 4038 \citep{Schirm2014}, NGC 2623, VV 114 \citep{Sliwa2013}, and Arp 220 \citep{Rangwala2011}. Dashed colored contours denote an advanced merger while a solid contour denotes an early/intermediate stage merger. Dash-dot lines represent log(Pressure) [cm$^{-3}$ K]. \label{fig:mergertempden}}
\end{figure*}

Figure \ref{fig:radexsolution} shows a clear difference in molecular gas conditions 
between NGC~2623, an advanced merger, and Arp~55, an early/intermediate merger. The 
molecular gas of NGC~2623 is warmer and less dense than that of Arp~55. We plot the most 
probable \tkin\ versus \nhtwo\ of Arp 220 \citep[][Sliwa $\&$ Downes 2016 submitted]{Rangwala2011}, Arp~299 \citep{Sliwa2012}, 
VV~114 \citep{Sliwa2013}, NGC~1614 \citep{Sliwa2014}, NGC~6240 \citep{Tunnard2015b}, NGC~4038/439 \citep{Schirm2014} and M51 (Schirm et al. 2016, submitted) along with NGC~2623 and Arp~55 (Figure \ref{fig:mergertempden}). We include M51 as a reference point. M51 is not a major merger but is an interacting system. For Arp~220 we have two points modeled at different scales (700 pc vs 1.5 kpc). The modelling of $Herschel$ data have constrained both a warm and cold component to the molecular gas \citep[e.g.][]{Rangwala2011,Schirm2014}. We note that our modelling is sensitive to only the cold component. Remarkably, there is a difference in physical conditions (on $\sim$ kiloparsec scales) between advanced mergers like Arp~220, NGC~2623 and VV~114 and early/intermediate mergers like Arp~299, NGC~1614, NGC~4038/39 + overlap and Arp~55: \textit{advanced mergers have a warmer and less dense molecular gas component than 
the early/intermediate mergers.} We note that the spatial resolutions of the observations vary 
for the sources we are comparing from $\sim$1-2~kpc, except for NGC~1614 where the spatial 
resolution was $\sim$ 230~pc. The physical conditions at several positions of NGC~1614 
revealed very similar results and once averaged at $\sim$ 1-2~kpc scales should yield a similar 
result. 

In the early phases of the merger process, the molecular gas is funnelled towards the inner 
kiloparsec regions \citep[e.g.][]{Hopkins2006,Kewley2006,Ellison2008}, which should drive the average volume density up. The 
starburst and/or AGN will warm the molecular gas as the merger progresses. In addition 
to heating the molecular gas, the massive stars will produce stellar winds that may dissipate 
some of the molecular gas and thus decrease the volume density. The supernova events 
produced by the massive stars can also play a role in decreasing the volume density. During the final coalescence, there should be another rapid increase in volume density as the gas mass of two disks merge into one. Arp 220 is very likely on this path.

In Figure \ref{fig:mergertempden}, we also see that there is a difference in thermal pressure between 
advanced mergers and early/intermediate stage mergers. The early/intermediate mergers 
have nearly an order of magnitude higher pressure than the advanced mergers. Work has 
been done in analyzing the pressure observationally and theoretically in disk galaxies over the 
last two decades \citep[e.g.][]{Elmegreen1993,Blitz2004,Blitz2006,Leroy2008}. 
\cite{Elmegreen1989} suggests that supernovae and other local activity affect the local average interstellar pressure by varying the scale height and therefore, the column density of stars that lies within the gas. This may be a likely scenario as the advanced mergers have had more time to form several generations of massive stars with stellar winds that have gone supernova.

\floattable
\begin{deluxetable*}{cccccccccc} %%%%Table 3
\tablecolumns{8} %necessary for multi-column lines
%\tablewidth{0pt}
\tabletypesize{\scriptsize}
%\rotate
\tablecaption{Bayesian RADEX Modeling Results \label{tab:results}}
\tablehead{ \colhead{Source} & \colhead{} &  \colhead{\tkin} &  \colhead{log(\nhtwo)} & \colhead{log(P)}  & \colhead{log(\ff)} & \colhead{log($<$N$_{^{12}CO}>$)} & \colhead{\xco}&\colhead{log(M$_{H_{2}}$)}& \colhead{\alphaco\tablenotemark{ab}}  \\
\colhead{} & \colhead{} &  \colhead{(K)} &  \colhead{(cm$^{-3}$)} & \colhead{(K cm$^{-3}$)}  &\colhead{} & \colhead{(cm$^{-2}$)} & \colhead{}& \colhead{(\msol)} &\colhead{(\alphacou)}   }    
\startdata
\hline
Arp~55NE & Mean & 54 & 3.6 & 5.4 & -0.2 & 18.0 & 32 & 8.3  & 0.2	\\
	& $\pm$1$\sigma$ range&11 -- 250 & 2.7 --4.6 & 4.8 -- 6.0 & -0.4 -- 0.1 & 17.8 -- 18.2 & 22 -- 50 & 8.1 -- 8.4	& 0.1 -- 0.2	\\
	& Best Fit& 19 & 3.9 & 5.1 & 0 & 17.9 & 24 & 8.2 &   	0.1		\\
\hline
Arp~55SW &Median & 36&4.0&5.5&-0.3 & 17.8 &20 &8.1  & 0.1		\\
	& $\pm$1$\sigma$ range&8 -- 175 & 2.8 -- 5.1 & 4.8 -- 6.2 & -0.5 -- 0.1	& 17.7 -- 18.0 & 14- 30 &8.0 -- 8.2 & 0.08 -- 0.1	\\
	& Best Fit& 13& 4.1 & 5.2 & -0.1 & 17.8 & 	15 &8.0 &	 0.08 	\\
\hline
NGC~2623 &Median & 125 & 2.8 & 4.8 &-1.0 &18.9 & 250 &8.7 & 0.5 \\
	& $\pm$1$\sigma$ range&31 -- 530&2.2 -- 3.3 & 4.5 -- 5.3 &-1.3 -- -0.7 & 18.4 -- 19.4 &  85 -- 720 &8.3 -- 9.2 &0.2 -- 1.7 \\
	& Best Fit& 107 & 2.7 & 4.8 & -1.2 &18.9 & 267 &8.8 & 0.7\\
\enddata
\tablenotetext{a}{\alphaco\ = 1.36\mmol\ /$L_{\rm{CO}}$. The factor of 1.36 is to account for He. }
\tablenotetext{b}{Assuming \xh\ = 3 $\times$ 10$^{-4}$}
\end{deluxetable*}

\section{Conclusions}\label{sec:conclusion}
In this paper, we have presented new ALMA and CARMA observations for Arp 55 and NGC 2623. We use the new 
data combined with existing SMA data to perform a non-LTE analysis to constrain the 
molecular gas physical conditions at about 1-1.5~kpc scales. 

\begin{enumerate}
\item We detect \tcoone\  and $J$ = 2-1 in both Arp 55 and NGC 2623 for the first time. 

\item We detect a CO blob in the tidal tail of Arp 55 that has active star formation. This blob has a 
mass of \mmol\ = 5.2 ($\frac{\alpha_{\rm{CO}}}{0.8}$) $\times$ 10$^{8}$ \msol\ and may be forming a tidal dwarf galaxy. More sensitive high-resolution observations of other tracers are needed to confirm this. 

\item The molecular gas in both nuclei of Arp 55 is cold ($\sim$ 10~K) and dense ($>$ 10$^{3}$ cm
$^{-3}$) while the molecular gas in NGC 2623 is warm ($\sim$ 110~K) and moderately dense 
(10$^{2.7}$ cm$^{-3}$). 

\item The \xco\ abundance ratio in NGC 2623 is unusually high ($>$ 200) when compared to the 
center of the Milky Galaxy while Arp 55 shows a more ``normal" ratio of $\sim$ 15-30. The \xco\ value found for 
NGC 2623 follows a trend that has been found in other LIRGs such as VV 114 \citep{Sliwa2013}, 
NGC 1614 \citep{Sliwa2014}, NGC 6240 \citep{Tunnard2015b}, Arp 193 \citep{Papadopoulos2014} and 
Mrk 231 \citep{Henkel2014}. The low \xco\ value for Arp 55 may indicate that the inflow of molecular gas has not yet enhanced \xco\ and/or nucleosynthesis has yet to produce enough massive stars that have enriched the ISM in $^{12}$C and thus \co\ and/or fractionation plays a larger role in determining the [\co]/[\tco] abundance ratio due to the cold environment.  

\item The \alphaco\ conversion factor is measured to be $\sim$ 0.1 and 0.7 ($\frac{3 \times 10^{-4}}
{x_{\rm{CO}}}$) \alphacou\ for Arp 55 and NGC 2623, respectively. To catch the transition from the Galactic-
like \alphaco\ to the LIRG-like value, we may need to look at an even younger merger than Arp 55 
such as Arp 240 \citep{Privon2013}. 

\item We find that, in general, advanced mergers such as NGC 2623 and Arp 220 have a molecular 
gas component that is warmer and less dense than early/intermediate mergers. There is also a 
pressure difference with advanced mergers having a lower pressure than early/intermediate stage 
mergers. Plausible scenarios to explain the pressure difference include mechanisms that can push 
back on the molecular gas like supernovae and stellar winds or the greater consumption of H$_{2}$ 
gas compared to HI. More work on the pressure in U/LIRGs is required. 

\end{enumerate}

\acknowledgments
We thank the anonymous referee for their comments and suggestions.
We thank T. Hattori for giving us the H$\alpha$ map and N. Rangwala and M.R.P Schirm for giving us the probability distributions for Arp~220 and the Antennae. KS thanks J. Kamenetzky for her help with the likelihood code. KS also thanks S. Aalto, D. Downes, T.J. Parkin, G.C. Privon, E. Schinnerer and M.R.P. Schirm for stimulating conversations and discussions. 
This paper makes use of the following ALMA data: ADS/JAO.ALMA$\#$2015.0.0804.S . ALMA is a partnership of ESO (representing its member states), NSF (USA) and NINS (Japan), together with NRC (Canada), NSC and ASIAA (Taiwan), and KASI (Republic of Korea), in cooperation with the Republic of Chile. The Joint ALMA Observatory is operated by ESO, AUI/NRAO and NAOJ.The National Radio Astronomy Observatory is a facility of the National Science Foundation operated under cooperative agreement by Associated Universities, Inc.

The Submillimeter Array is a joint project between the Smithsonian 
Astrophysical Observatory and the Academia Sinica Institute of Astronomy and Astrophysics and is 
funded by the Smithsonian Institution and the Academia Sinica. 

The James Clerk Maxwell Telescope has historically been operated by the Joint Astronomy Centre on behalf of the Science and Technology Facilities Council of the United Kingdom, the National Research Council of Canada and the Netherlands Organisation for Scientific Research.

Support for CARMA construction was derived from the states of California, Illinois, and Maryland, the James S. 
McDonnell Foundation, the Gordon and Betty Moore Foundation, the Kenneth T. and Eileen L. 
Norris Foundation, the University of Chicago, the Caltech Associates, and the NSF. 
Historically, ongoing CARMA development and operations were supported by the NSF under a cooperative agreement, and by the CARMA partner universities. 

C.D.W. acknowledges support by the Natural Science and Engineering Research Council of Canada (NSERC). K.S. acknowledges support by the Ontario Graduate Scholarship (OGS). This research made use of APLpy, an open-source plotting package for Python hosted at http://aplpy.github.com. This research made use of the python plotting package matplotlib \citep{Hunter2007}.


%% for the paper.  Each keyword is check against the master list during
%% copy editing.  Individual instruments can be provided in parentheses,
%% after the keyword, but they are not verified.

\vspace{5mm}
\facilities{ALMA, CARMA, JCMT, SMA}

\software{CASA, Python (matplotlib, APLpy, astropy), RADEX, starlink}

\appendix

\section{Line Ratios} \label{sec:lineratio}%%%%%%%%%LINE RATIOS

\begin{figure*}[htbp] %%%FIGURE 3 Arp 55 line ratio maps
\centering
\gridline{\fig{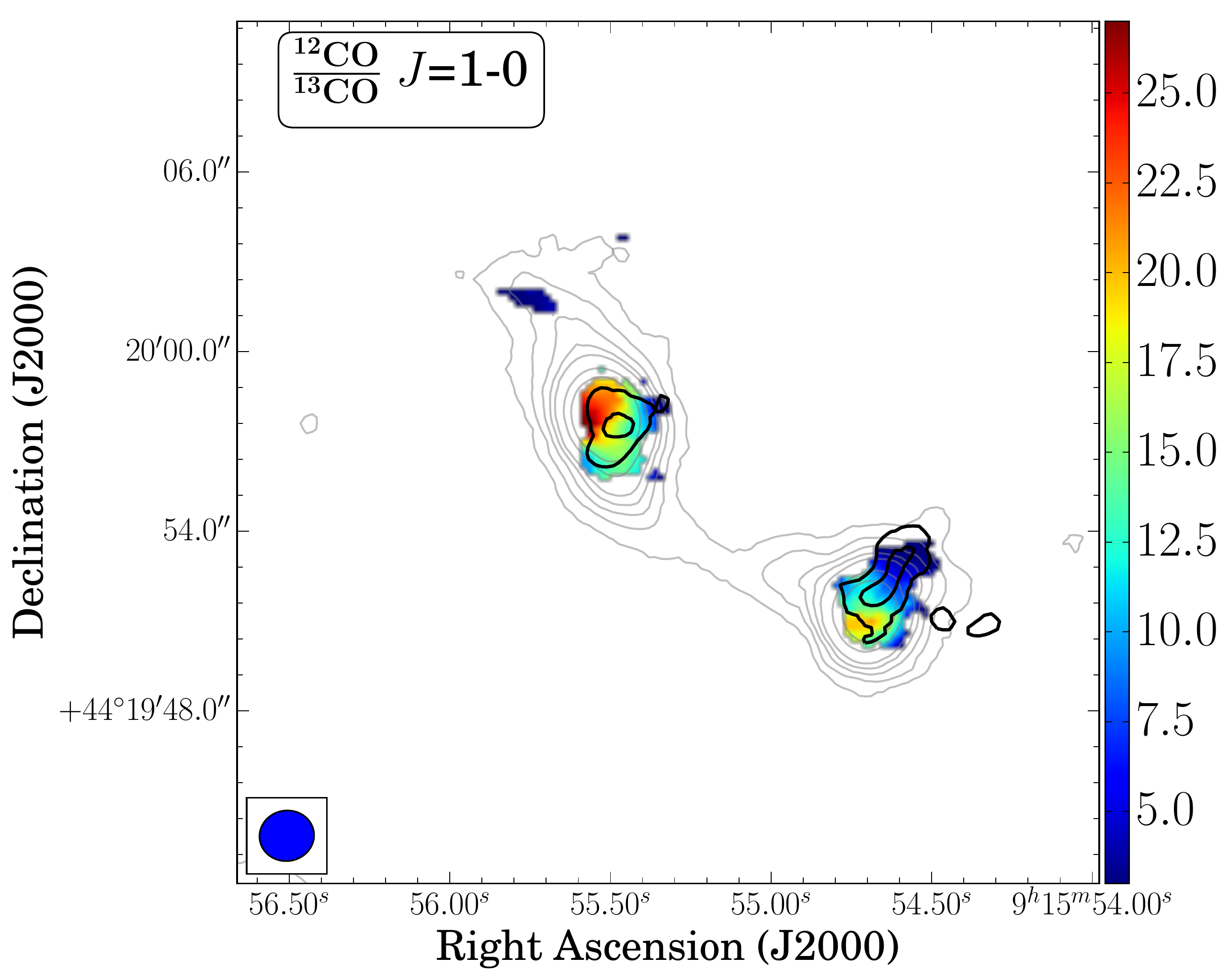}{0.4\textwidth}{(a)}
\fig{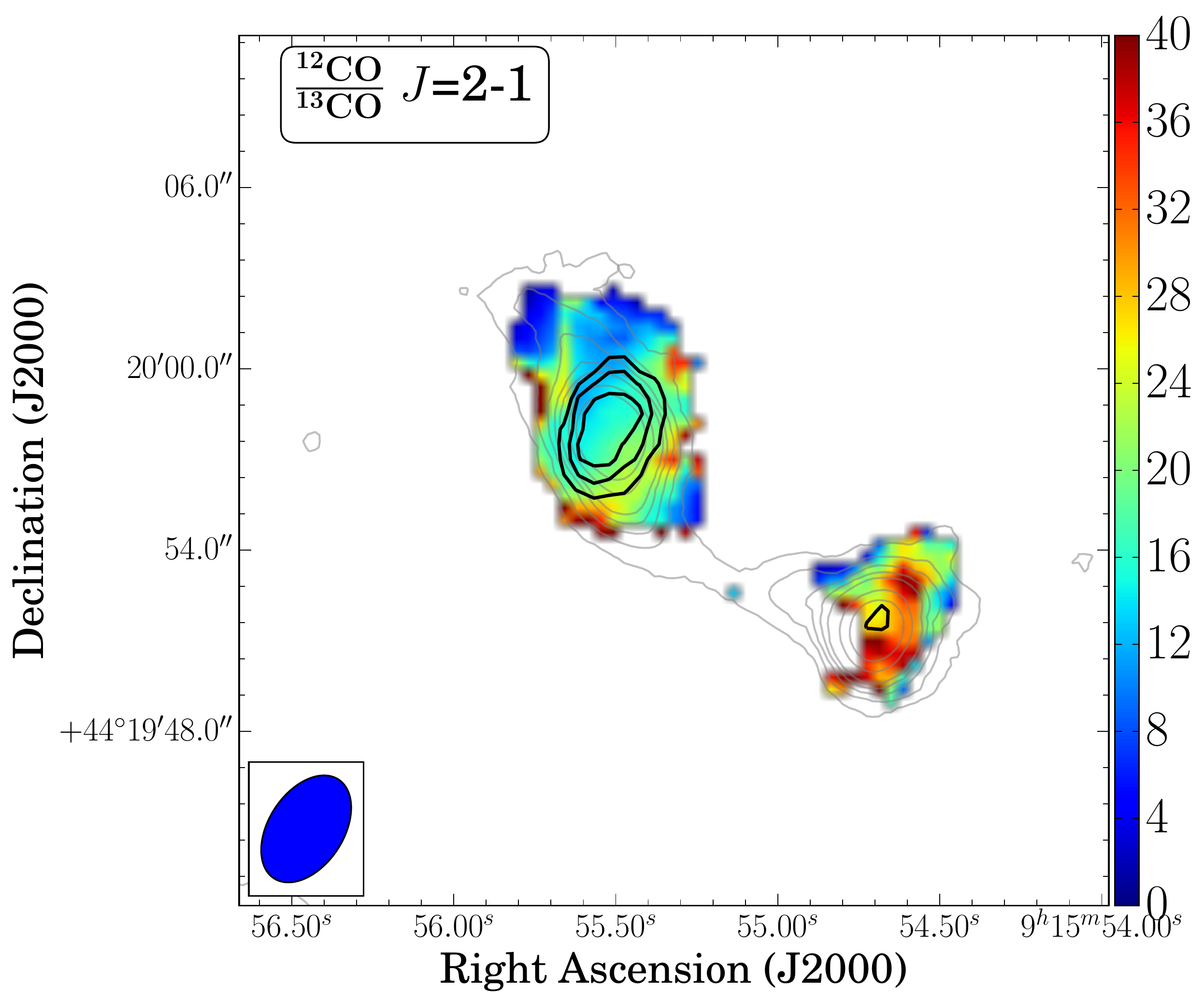}{0.4\textwidth}{(b)}}
\gridline{\fig{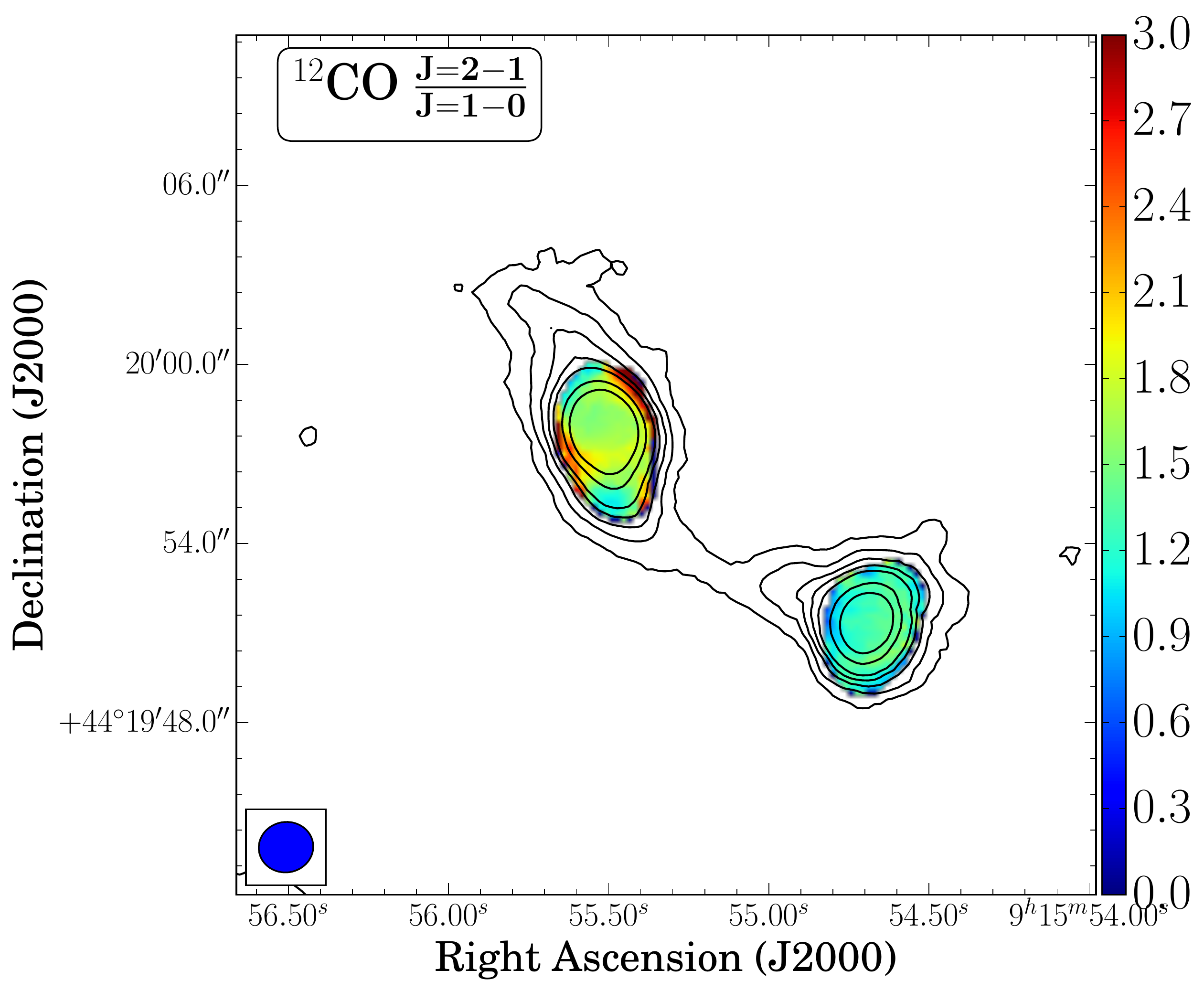}{0.4\textwidth}{(c)}
\fig{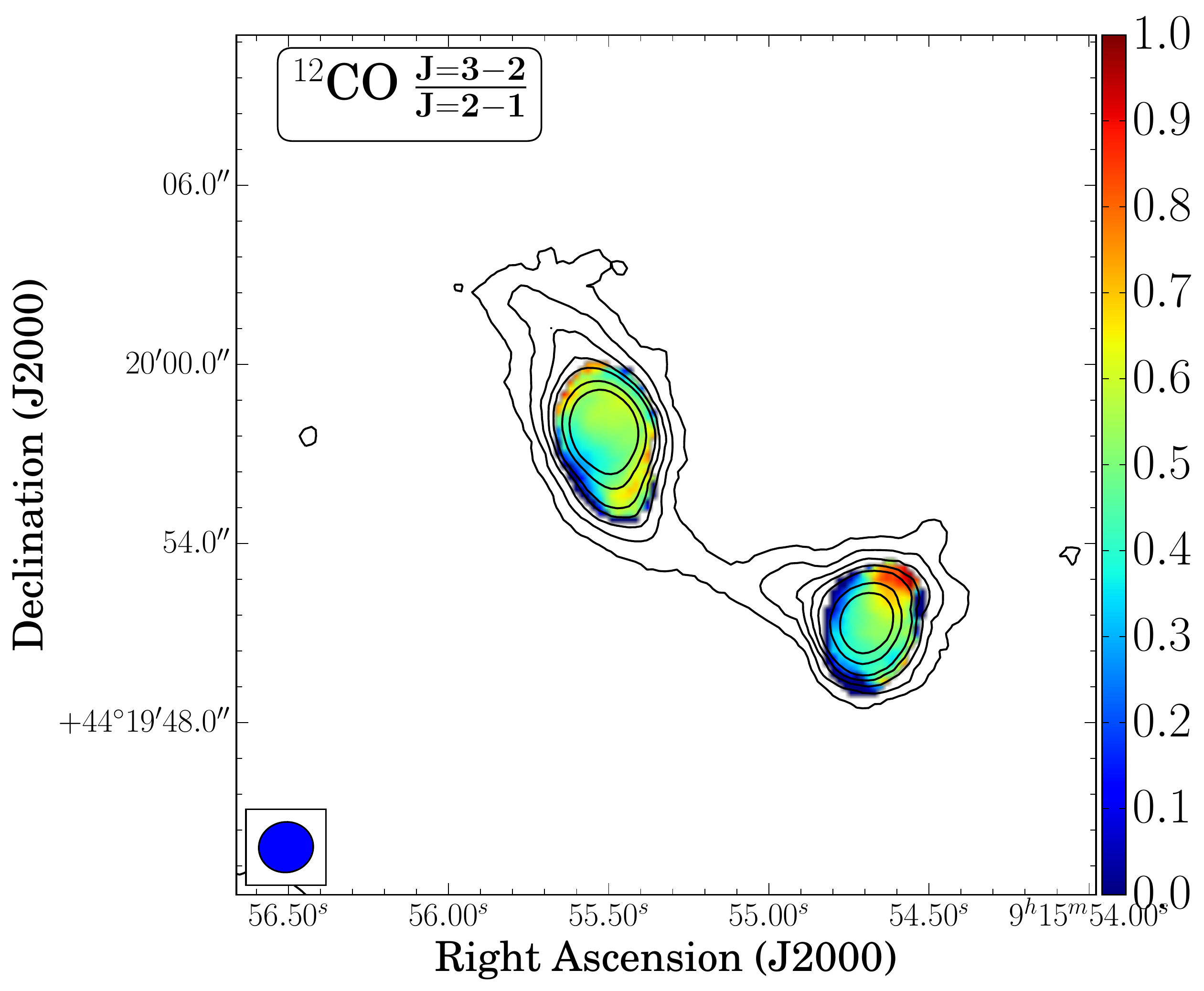}{0.4\textwidth}{(d)}}
\fig{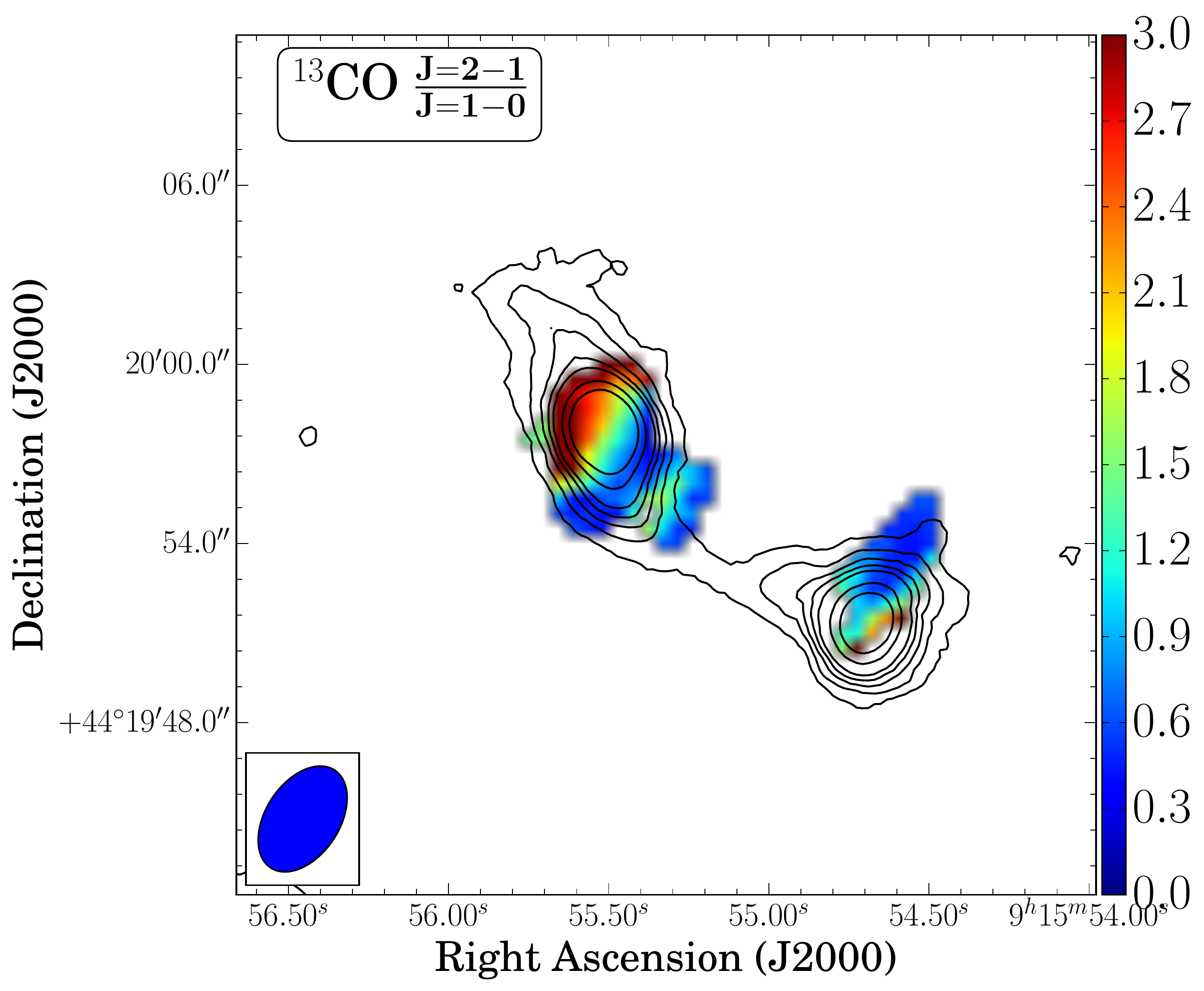}{0.4\textwidth}{(e)}
\caption{Integrated brightness temperature line ratio maps for Arp 55: (a) $\frac{I_{\rm{12CO(1-0)}}}{I_{\rm{13CO(1-0)}}}$ ($R_{10}$), (b) $\frac{I_{\rm{12CO(2-1)}}}{I_{\rm{13CO(2-1)}}}$ (R$_{21}$) (c)  $\frac{I_{\rm{12CO(2-1)}}}{I_{\rm{12CO(1-0)}}}$  ($r_{21}$), (d) $\frac{I_{\rm{12CO(3-2)}}}{I_{\rm{12CO(2-1)}}}$ (r$_{32}$) and (e) $\frac{I_{\rm{13CO(2-1)}}}{I_{\rm{13CO(1-0)}}}$ ($^{13}r_{21}$). Contours are of \coone\ for reference. The additional contours in panels (a) and (b) denote SN ratios of [3, 4] and [2, 3, 4], respectively. The ellipse in the bottom left corner of each map represents the synthesized beam. \label{fig:arp55lineratios}}
\end{figure*}

\begin{figure*}[htbp] %%%FIGURE 4 NGC 2623 line rati o maps
\centering
\gridline{\fig{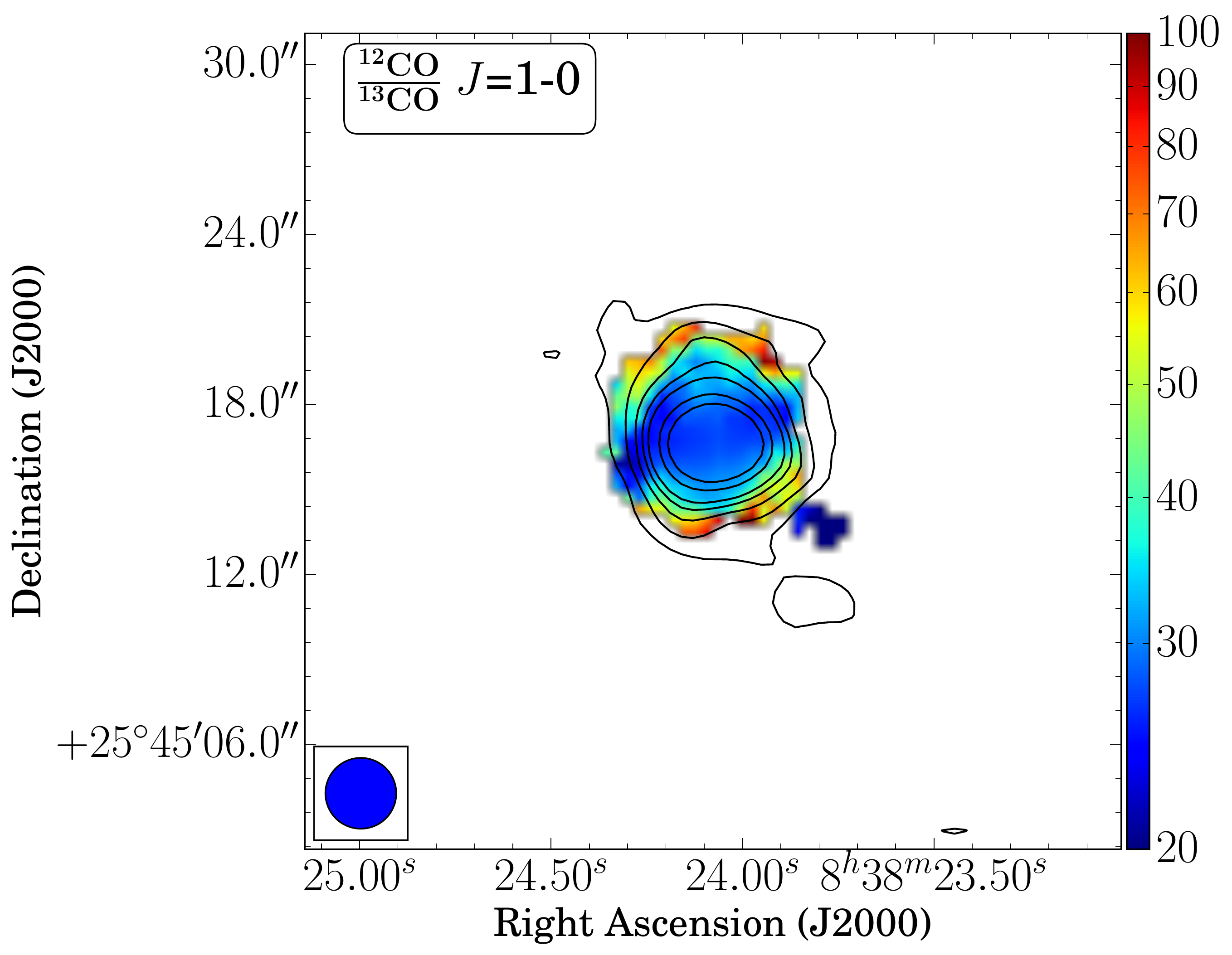}{0.4\textwidth}{(a)}
\fig{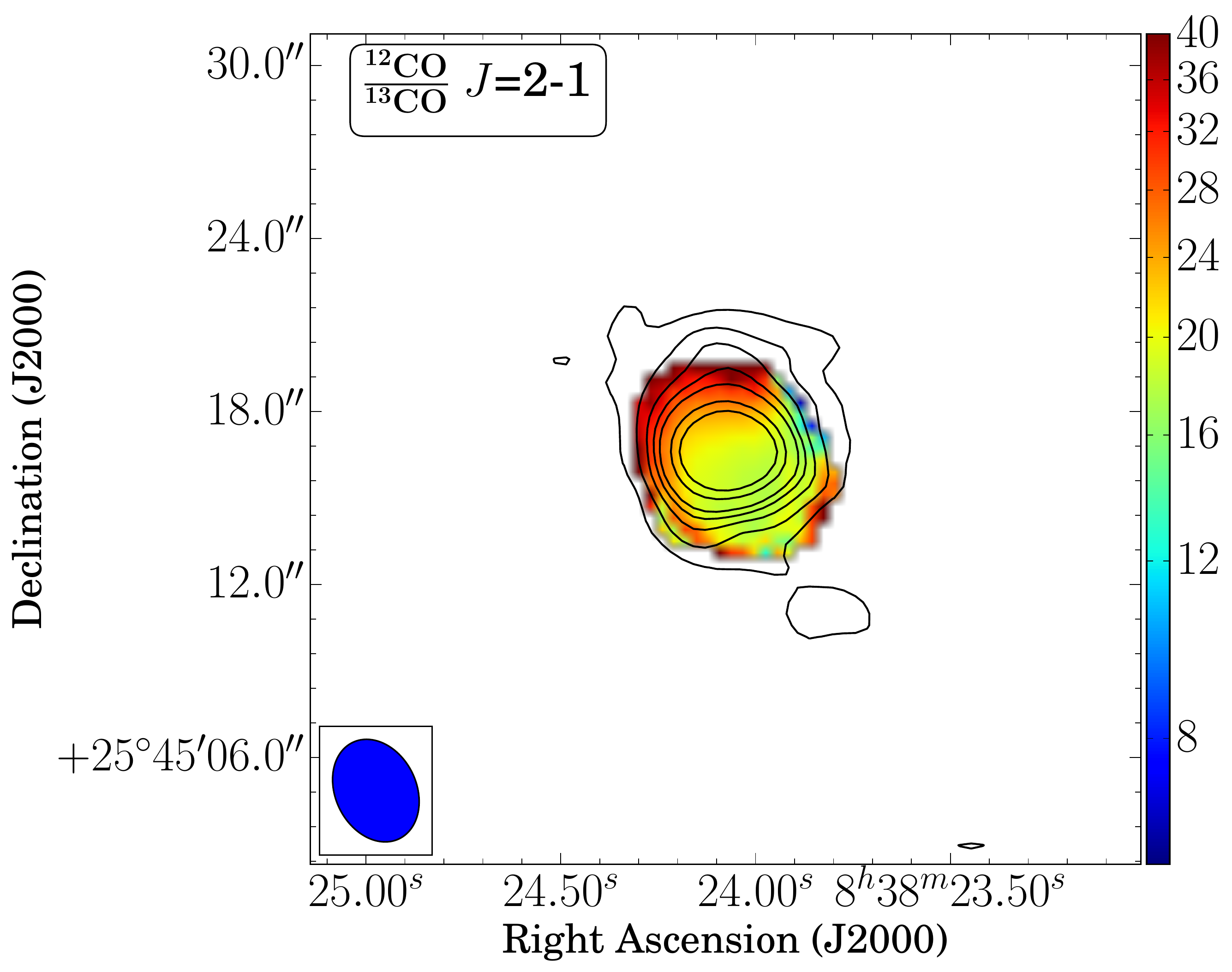}{0.4\textwidth}{(b)}}
\gridline{\fig{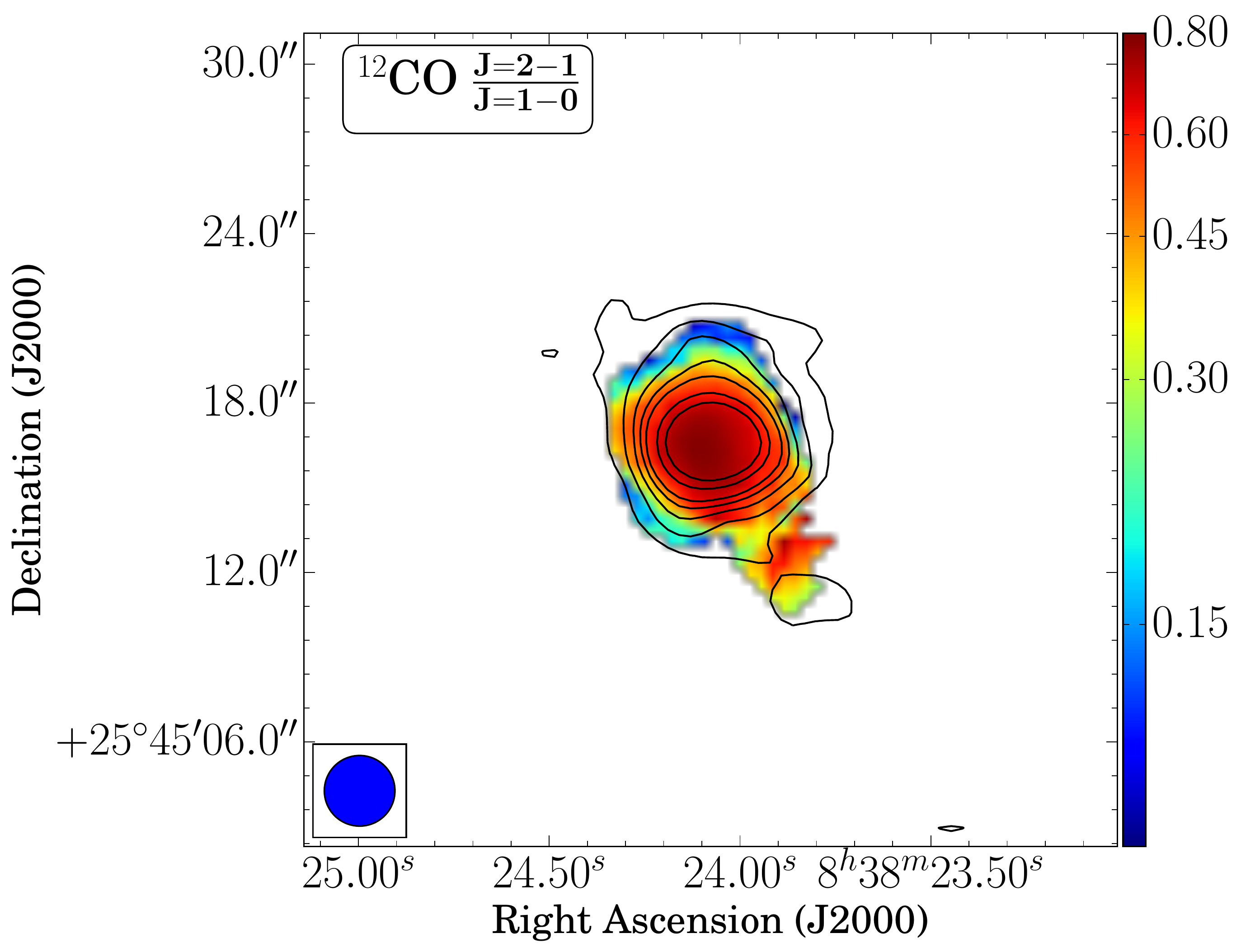}{0.4\textwidth}{(c)}
\fig{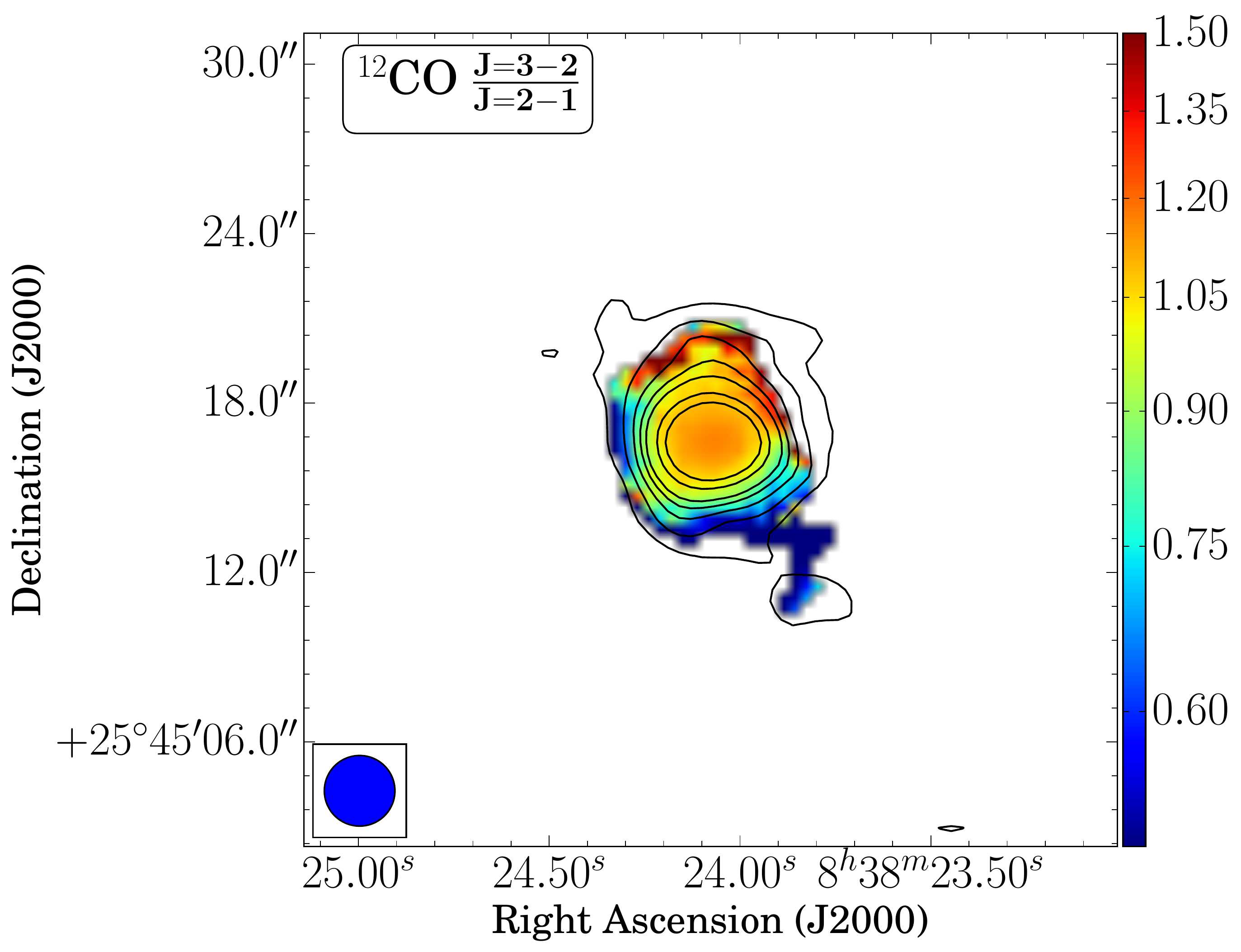}{0.4\textwidth}{(d)}}
\fig{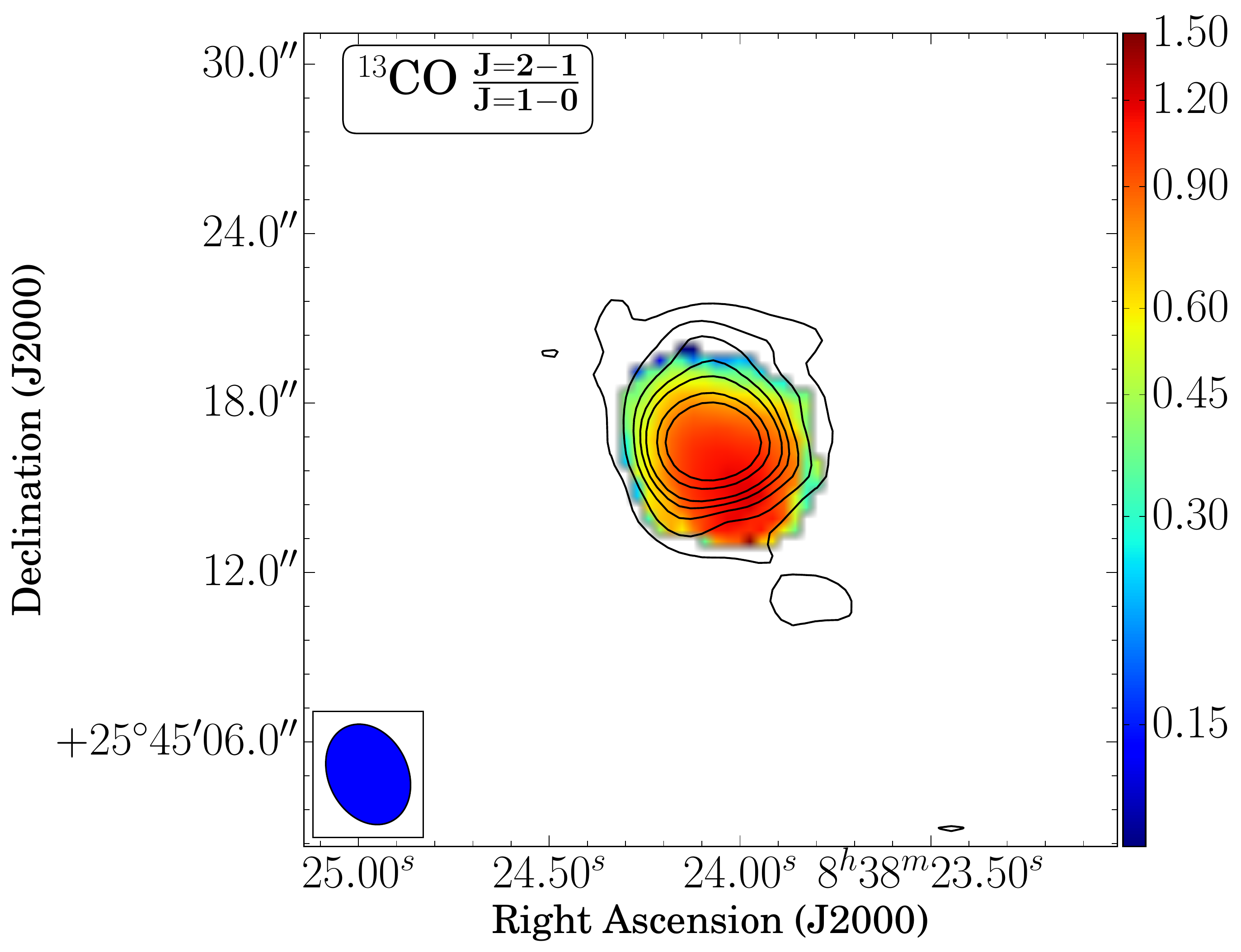}{0.4\textwidth}{(e)}

\caption{Integrated brightness temperature line ratio maps for NGC 2623: : (a) $\frac{I_{\rm{12CO(1-0)}}}{I_{\rm{13CO(1-0)}}}$ ($R_{10}$), (b) $\frac{I_{\rm{12CO(2-1)}}}{I_{\rm{13CO(2-1)}}}$ (R$_{21}$) (c)  $\frac{I_{\rm{12CO(2-1)}}}{I_{\rm{12CO(1-0)}}}$  ($r_{21}$), (d) $\frac{I_{\rm{12CO(3-2)}}}{I_{\rm{12CO(2-1)}}}$ (r$_{32}$) and (e) $\frac{I_{\rm{13CO(2-1)}}}{I_{\rm{13CO(1-0)}}}$ ($^{13}r_{21}$). Contours are of \coone\ for reference. The ellipse in the bottom left corner of each map represents the synthesized beam.  \label{fig:n2623lineratios}}
\end{figure*}

Line ratios can be used to estimate excitation conditions of the molecular gas in a galaxy. 
We create the following integrated brightness temperature (I = $\int$T$_{B}$dV) line ratios: \\
$r_{21}$ = $\frac{I_{\rm{12CO(2-1)}}}{I_{\rm{12CO(1-0)}}}$,\\
$r_{32}$ = $\frac{I_{\rm{12CO(3-2)}}}{I_{\rm{12CO(2-1)}}}$,\\
$R_{10}$ =$\frac{I_{\rm{12CO(1-0)}}}{I_{\rm{13CO(1-0)}}}$,\\
$R_{21}$ =$\frac{I_{\rm{12CO(2-1)}}}{I_{\rm{13CO(2-1)}}}$ and \\
$^{13}r_{21}$ = $\frac{I_{\rm{13CO(2-1)}}}{I_{\rm{13CO(1-0)}}}$  (Figures \ref{fig:arp55lineratios}, \ref{fig:n2623lineratios}).

For each line ratio map, we match the angular resolution using a tapering weight during imaging. For Arp~55 we degrade the resolution to that of the \tcoone\ map (see Table \ref{tab:summary}) and for NGC~2623 we degrade the resolution to 2.5\arcsec.  For the 
$^{13}r_{21}$ and $R_{21}$ line ratio maps, we degrade the \tcoone\ and \cotwo\ map 
resolutions to match the \tcotwo\ resolution for both sources..  We also make a 1$\sigma$ and 2$\sigma$ cutoff in the \tco\ and \co\  
integrated intensity maps, respectively, when creating the line ratio maps (Figures \ref{fig:arp55lineratios} and \ref{fig:n2623lineratios}).

Since the signal-to-noise (SN) ratio for the \tco\ observations of Arp~55 is low($<$6), we apply an additional cutoff to the corresponding \co\ map in order to match the SN ratio (8$\sigma$ and 7$\sigma$ for \coone\ and $J$=2-1, respectively).  For R$_{10}$, the majority of the emission lies within regions of  $\geq$3$\sigma$ detections in \tco. Towards the edges of the sources, the SN ratio drops and therefore, the uncertainty approaches 50$\%$. We can therefore treat edges as upper limits. For R$_{21}$, a significant number of pixels above the initial cutoff lie in regions below 2$\sigma$ \tco\ emission and should be perceived as upper limits. We note that this is severe for Arp~55SW and the variation in the R$_{21}$ line ratio should be taken with caution. 

The $r_{21}$ and $r_{32}$ line ratios for both Arp 55 and NGC 2623 vary smoothly
indicating no extreme changes in the conditions of the molecular gas. The $r_{21}$ line ratio 
for Arp 55 is $>$1 for both nuclei which may suggest that \coone\ is optically thin. Indeed, this is consistent with the RADEX solutions for the best fit where $\tau_{1-0}$ = 0.2 - 0.3 for Arp~55. We note 
that similar $r_{21}$ line ratios were found for Arp 299 \citep{Sliwa2012}, another early/
intermediate merger stage LIRG.

%%%%%%%%%%%%%%

%%%%%%%%%%%%%

\begin{deluxetable*}{cccccccc} %TABLE  LINE RATIOs
\tablewidth{0pt}
\tablecaption{Line Ratios \label{tab:lineratios}}
\tablehead{
\colhead{Source} &\colhead{$\alpha_{\rm{J2000}}$}&\colhead{$\delta_{\rm{J2000}}$}& \colhead{$r_{21}$} &\colhead{$r_{32}$} &
\colhead{$R_{10}$} &\colhead{$R_{21}$} & \colhead{$^{13}r_{21}$}  } 
\startdata
Arp 55 NE	&	09$^{\rm{h}}$15$^{\rm{m}}$55$^{\rm{s}}$.5	& +44$^{\circ}$19$^{\prime}$57\arcsec.9	&1.5 $\pm$ 0.4	&0.4 $\pm$ 0.1	&22 $\pm$ 5	&19 $\pm$ 4 &2.1 $\pm$ 0.9	   \\
Arp 55 SW&	09$^{\rm{h}}$15$^{\rm{m}}$54$^{\rm{s}}$.7	& +44$^{\circ}$19$^{\prime}$51\arcsec.5	&1.3 $\pm$ 0.4	&0.4 $\pm$ 0.1	&13 $\pm$ 3	& \textbf{$<$25 $\pm$ 6} &1.6 $\pm$ 0.6	   \\
NGC 2623&	08$^{\rm{h}}$38$^{\rm{m}}$24$^{\rm{s}}$.1	&+25$^{\circ}$45$^{\prime}$16\arcsec.53	&0.8 $\pm$ 0.2	& 1.2 $\pm$ 0.3	&27.1 $\pm$ 4.9	& 20.1 $\pm$ 5.0  & 1.0 $ \pm$ 0.2
\enddata
%\tablenotetext{}{Note: Line ratios measured from integrated brightness temperature maps}
\end{deluxetable*}

\section{Velocity Maps}\label{sec:velmaps}
\begin{figure*}[htbp] %%%FIGURE 5 NGC 2623 line ratio maps

\gridline{\fig{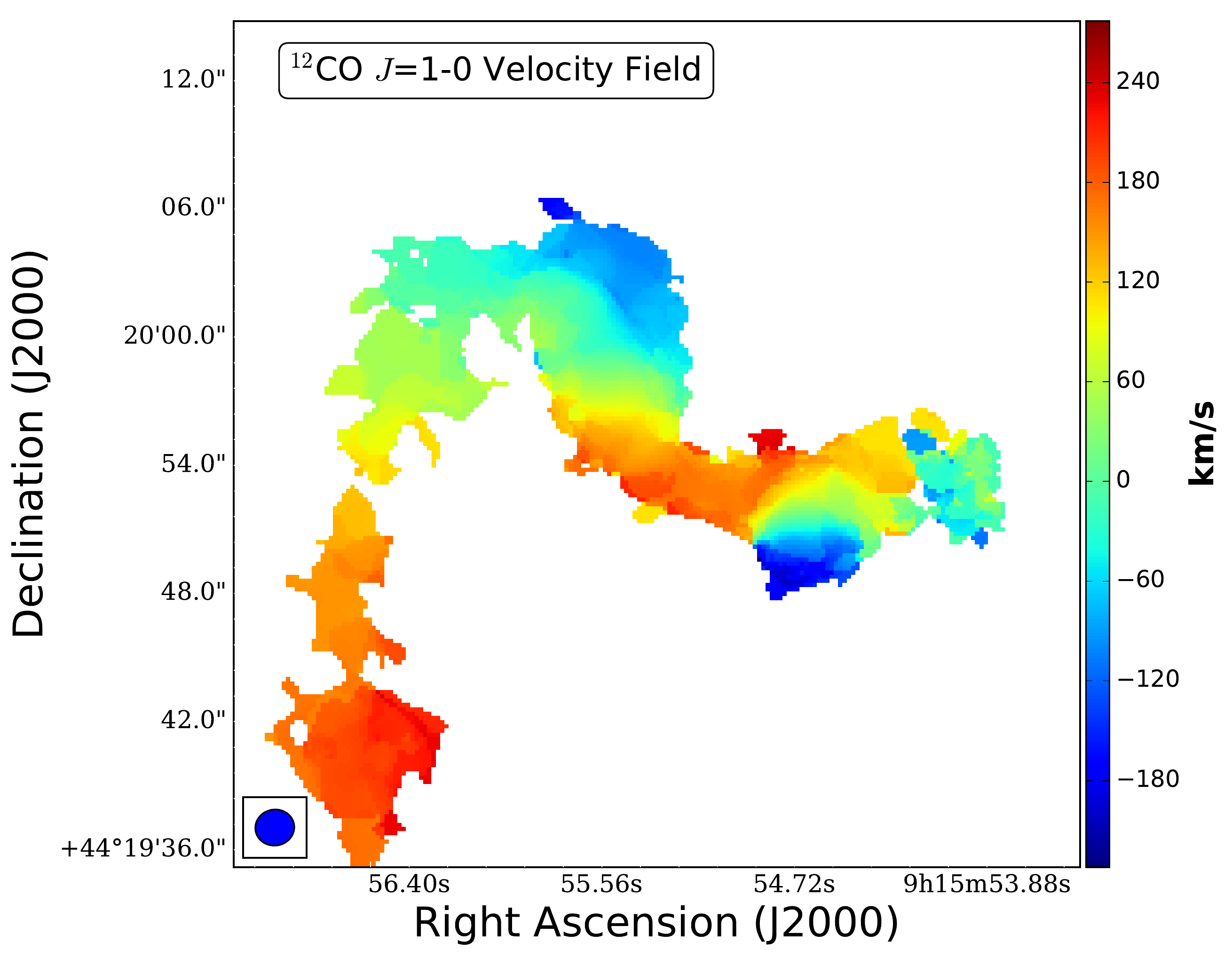}{0.5\textwidth}{(a)}
\fig{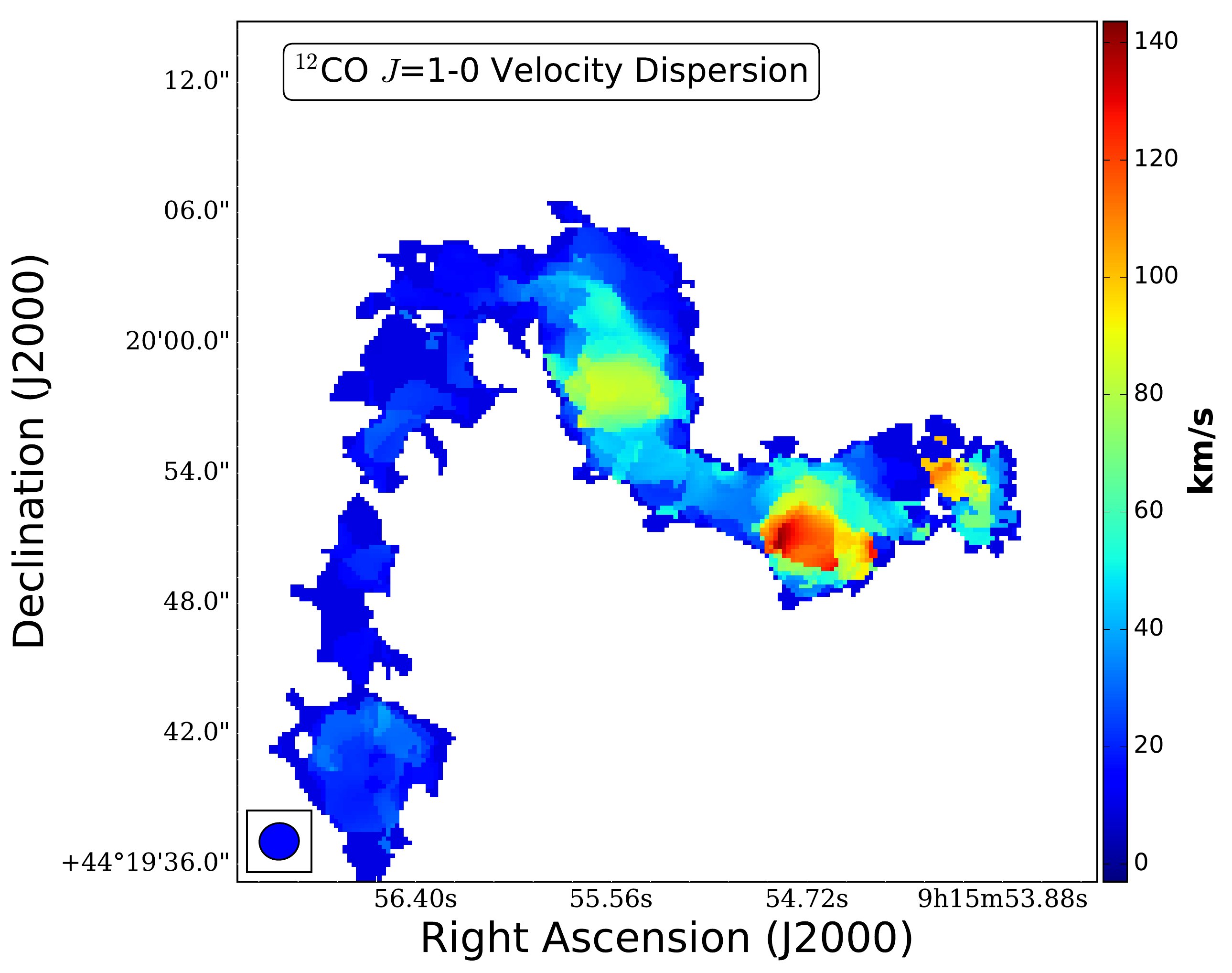}{0.5\textwidth}{(b)}}

\gridline{\fig{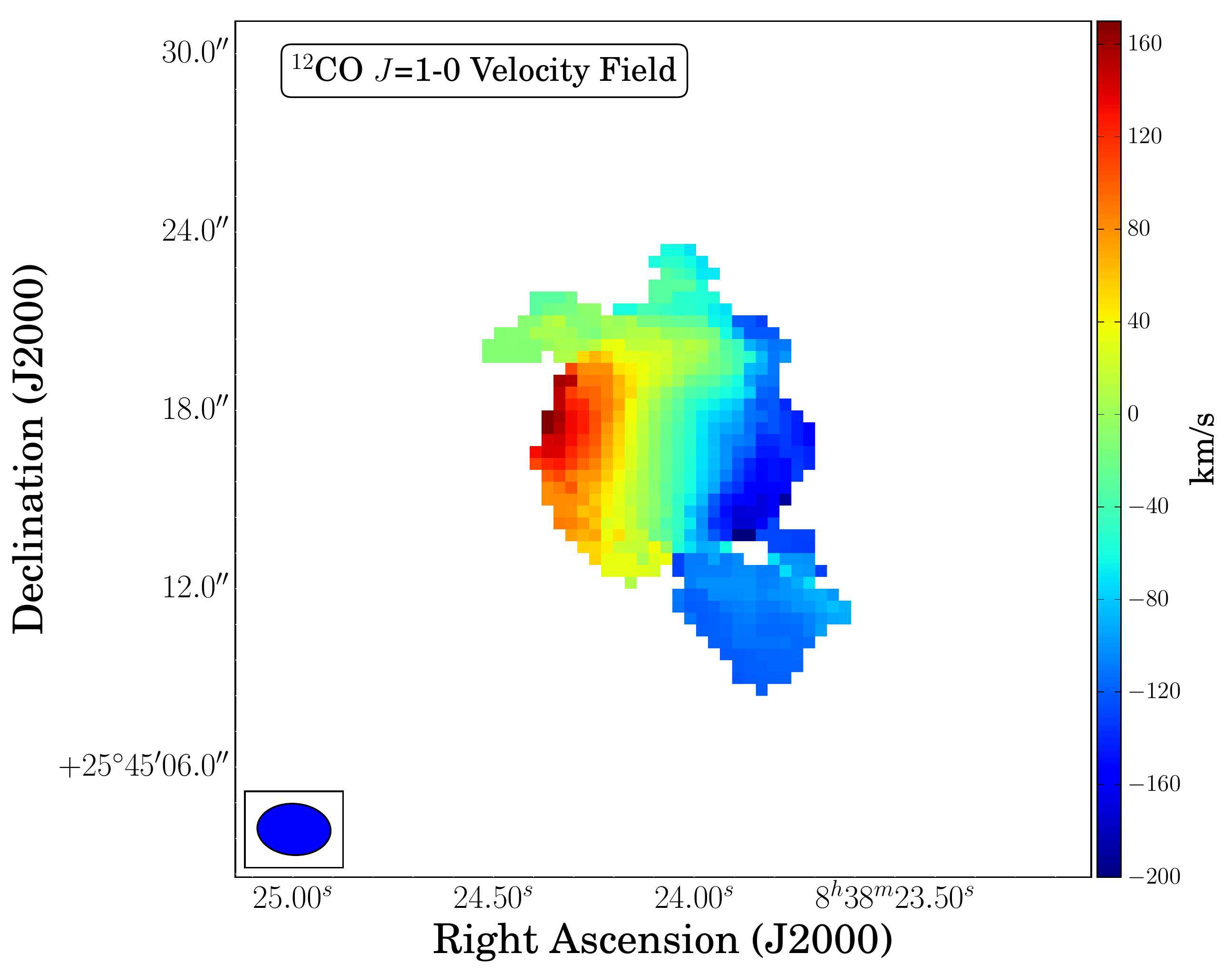}{0.5\textwidth}{(c)}
\fig{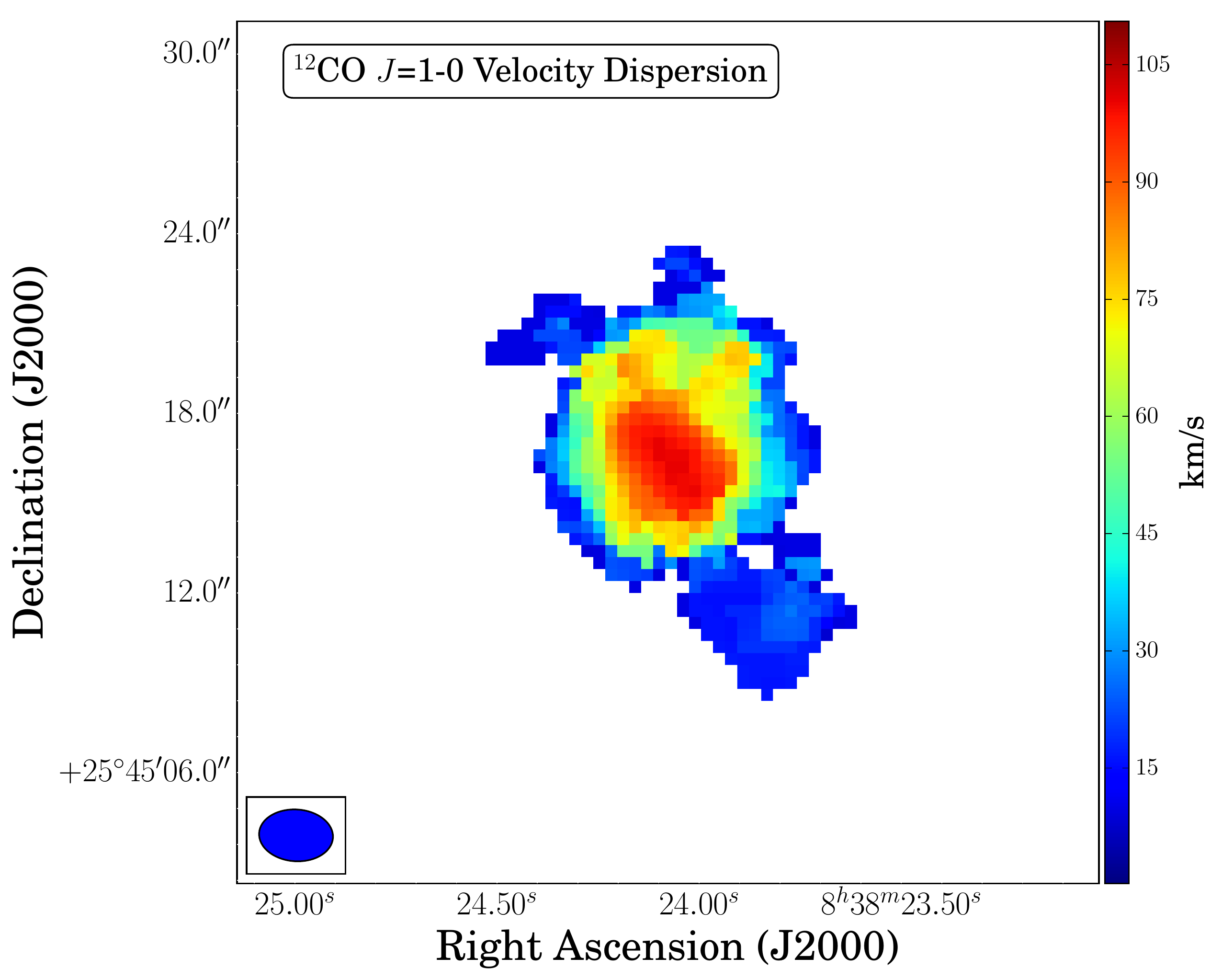}{0.5\textwidth}{(d)}}
\caption{CARMA \coone\ velocity field and dispersion maps of (a, b) Arp 55  and (c, d) NGC 2623. \label{fig:velmaps}}
\end{figure*}

Using the new CARMA \coone\ maps, we create velocity field (moment 1) and dispersion (moment 2; $\sigma$ = FWHM/2$\sqrt{2\rm{ln}(2)}$) maps. We only include channels that contain emission above 2$\sigma$. The maps are presented in Figure \ref{fig:velmaps}. 

%\bibliography{../../master-reference-list.bib}
%\bibliographystyle{aasjournal.bst}



\end{document}